\documentclass[12pt]{article}
\usepackage{amsmath, amssymb, amsthm, mathrsfs, latexsym,float,psfrag,epsfig,natbib,comment}
\usepackage{multirow,booktabs,graphicx,color,epstopdf,enumerate,threeparttable,hyperref}

\newcommand{\blind}{0}

\floatstyle{ruled}
\newfloat{algorithm}{tbhp}{loa}
\floatname{algorithm}{Algorithm}

\theoremstyle{plain}

\newtheorem{algo}{Algorithm}

\newtheorem{proposition}{Proposition}

\def\be{\begin{equation}}
\def\ee{\end{equation}}
\def\ben{\begin{equation*}}
\def\een{\end{equation*}}
\def\bea{\begin{eqnarray}}
\def\eea{\end{eqnarray}}
\def\bean{\begin{eqnarray*}}
\def\eean{\end{eqnarray*}}

\def\diag{\mathrm{diag}}
\def\cL{\mathcal{L}}
\def\cD{\mathcal{D}}
\def\cP{\mathcal{P}}
\def\cS{\mathcal{S}}
\def\cN{\mathcal{N}}
\def\bx{\mathbf{x}}
\def\ba{\mathbf{a}}
\def\cD{\mathcal{D}}
\def\cDD{\mathcal{DD}}
\def\cSDD{\mathcal{SDD}}
\def\cI{\mathcal{I}}
\def\cM{\mathcal{M}}

\newcommand{\bs}[1]{\boldsymbol{#1}}

\begin{document}

\def\spacingset#1{\renewcommand{\baselinestretch}%
{#1}\small\normalsize} \spacingset{1}


\if0\blind {
  \title{\Large \bf Diagonally-Dominant Principal Component Analysis\thanks{The authors gratefully acknowledge the
    support of NIH grant P50DA039838 and NSF grants DMS-1505256, DMS-1712958 and DMS–1811552.}}
  \author{Zheng Tracy Ke\hspace{.2cm}\\
    Department of Statistics, Harvard University\\
    and \\
    Lingzhou Xue\hspace{.2cm}\\
    Department of Statistics, Pennsylvania State University\\
    and \\
    Fan Yang\hspace{.2cm}\\
    Department of Statistics, University of Chicago}
  \maketitle
} \fi

\if1\blind {
  \bigskip
  \bigskip
  \bigskip
  \begin{center}
    {\LARGE\bf Low-rank and diagonally-dominant covariance matrix decomposition: methodology and applications}
\end{center}
  \medskip
} \fi

\bigskip
\begin{abstract}
We consider the problem of decomposing a large covariance matrix into the sum of a low-rank matrix and a diagonally dominant matrix, and we call this problem the ``Diagonally-Dominant Principal Component Analysis (DD-PCA)''. DD-PCA is an effective tool for designing statistical methods for strongly correlated data. We showcase the use of DD-PCA in two statistical problems: covariance matrix estimation, and global detection in multiple testing.
Using the output of DD-PCA, we propose a new estimator for estimating a large covariance matrix with factor structure. Thanks to a nice property of diagonally dominant matrices, this estimator enjoys the advantage of simultaneous good estimation of the covariance matrix and the precision matrix (by a plain inversion). A plug-in of this estimator to linear discriminant analysis and portfolio optimization yields appealing performance in real data. We also propose two new tests for testing the global null hypothesis in multiple testing when the $z$-scores have a factor covariance structure. Both tests first use DD-PCA to adjust the individual $p$-values and then plug in the adjusted $p$-values to the Higher Criticism (HC) test. These new tests significantly improve over the HC test and compare favorably with other existing tests. For computation of DD-PCA, we propose an iterative projection algorithm and an ADMM algorithm.
\end{abstract}

\noindent%
{\it Keywords:} ADMM, Alternating Projection, Approximate Factor Model, Covariance Matrix Estimation, De-Correlation, Higher Criticism, Global Testing, POET\vfill

\newpage
\spacingset{1} 

\section{Introduction} \label{sec:intro}
The {\it approximate low-rankness} is a popular structural assumption on covariance matrices. It assumes that a $p\times p$ covariance matrix $\bs{\Sigma}$ decomposes into
\begin{equation} \label{apprx-low-rank}
\bs{\Sigma} = \bs{L} + \bs{A}, \qquad \mbox{where}\quad \mathrm{rank}(\bs{L})=K\ll p, \quad \mbox{and $\bs{A}$ is a ``nice" matrix}.
\end{equation}
Equivalently, it introduces a latent factor model on any random vector $X$ whose covariance matrix is $\bs{\Sigma}$, where $\bs{A}$ is the ``residual covariance matrix" after the effects of latent variables are removed. Such a decomposition is not unique and varies with the meaning of a ``nice" $\bs{A}$. One can impose different requirements on $\bs{A}$ to facilitate different applications. In the classical factor models for econometrics and finance, $\bs{A}$ is assumed a diagonal matrix \citep{fama1993common} or a sparse matrix \citep{chamberlain-1983}, to enforce that the idiosyncratic noise accounts for little cross-sectional risk. In large-scale multiple testing, it is often assumed that the covariance matrix of test statistics has the above decomposition with $\bs{A}$ being a diagonal matrix  \citep{leek-storey-2008} or having a small Frobenius norm \citep{fan-han-gu-2012}. The motivation there is development of factor-adjusted multiple testing procedures, to make it legitimate to use conventional multiple testing methods on the post-factor-removal data. In image processing, a similar decomposition on image matrices was proposed \citep{rpca}, where $\bs{A}$ is assumed sparse, for the purpose of capturing details of images. In this paper, we explore a new type of {\it approximate low-rankness} where
\begin{equation} \label{assmp-A}
\mbox{Each diagonal of $\bs{A}$ is large compared with other entries in the same row}.
\end{equation}
Translated to the latent variable representation, it means, after the effects of latent variables are removed, the {\it correlation} matrix of ``residual" variables have uniformly small off-diagonal entries. One motivation of imposing this condition is to take into account the varying scale of the diagonal elements of $\bs{A}$. Most aforementioned approximate low-rank decompositions first perform PCA on $\bs{\Sigma}$ (or an empirical version of it) to remove the first a few principal components, and then conduct operations on the remaining matrix. It is often observed that the diagonal elements of the remaining matrix has considerable variations in magnitude. To deal with it requires careful adjustment on the post-PCA operations, such as adaptive thresholding \citep{cai2011adaptive}. On the contrary, we impose the requirement \eqref{assmp-A} directly in the decomposition \eqref{apprx-low-rank}, in hopes of improving the PCA factors and easing the post-PCA operations. Another motivation of adopting the assumption \eqref{assmp-A} is to guarantee that $\bs{A}^{-1}$ is well-behaved. In many applications such as portfolio management and linear discriminant analysis, $\bs{A}^{-1}$ plays a key role (see Section~\ref{sec:CovEst}). In the decomposition \eqref{apprx-low-rank}, forcing $\bs{A}$ to be a strictly diagonal matrix can ensure both $\bs{A}$ and $\bs{A}^{-1}$ are well-behaved, but this requirement is often too restrictive, and \eqref{assmp-A} is a natural relaxation. We note that imposing the common sparsity assumption on $\bs{A}$ does not even guarantee positive definiteness. Despite of remedies such as increasing the threshold or projection to the positive semi-definite cone \citep{fan2016overview}, these approaches still don't guarantee that $\bs{A}^{-1}$ is a ``nice" matrix.

To formulate \eqref{assmp-A} mathematically, we define the set of ``symmetric $c$-diagonally-dominant" matrices, for any $c>0$:
\be \label{SDD}
\cSDD_c^+ = \Bigl\{\bs{A} = (a_{ij})_{p\times
p}: \bs{A}^T=\bs{A},\; a_{jj} \geq c\sum_{i: i\not= j} |a_{ji}|\
\mbox{ for\ all\ }1\leq j\leq p \Bigr\}.
\ee
For $c=1$, it reduces to the usual definition of diagonally-dominant matrices, and we omit the subscript and write $\cSDD_1^+=\cSDD^{+}$. Given a $p\times p$ positive semi-definite matrix $\bs{S}$, we introduce an optimization problem:
\be\label{dd-PCA}
\min_{(\bs{L},\bs{A})} \
\|\bs{S} - \bs{L}-\bs{A} \|_F, \qquad \text{subject to} \quad
\mathrm{rank}(\bs{L})\leq K, \;\; \bs{L}=\bs{L}^T,\;\;  \bs{A}\in
\cSDD_c^+,
\ee
where $\|\cdot\|_F$ is the
matrix Frobenius norm. We call it {\it
Diagonally-Dominant Principal Component Analysis (DD-PCA)}. In this paper, we are primarily interested in $c=1$; discussions of $c\neq 1$ are deferred to Section~\ref{sec:methodology}.

The definition of DD-PCA is a nonconvex optimization with a rank constraint. Similar to solving other rank constrained optimizations in matrix completion, one can either solve a convex relaxation of \eqref{dd-PCA} or develop an iterative algorithm that converges to a local minimum of \eqref{dd-PCA}. These ideas generate several variants of DD-PCA, as detailed in Section~\ref{sec:methodology}. Among those variants, one is of particular interest, which we call
{\it One-step DD-PCA}:
\begin{itemize}
\item PCA: Obtain the $K$ leading eigenvalues and eigenvectors of $\bs{S}$, denoted as $\lambda_1\geq \ldots\geq \lambda_K\geq 0$ and $\xi_1,\ldots, \xi_K\in\mathbb{R}^p$. Let $\bs{L}=\sum_{k=1}^K\lambda_k\xi_k\xi_k^T$.
\item Projection to $\cSDD^+$: Initialize $\bs{A}^{(0)}=\bs{S}-\bs{L}$ and $\bs{J}^{(0)}=\bs{0}$. For $t=1,2,\ldots,$
\begin{itemize}
\item Run the MRT algorithm \citep{mendoza-etal-1998}\footnote{The MRT algorithm computes the unique projection of a $p\times p$ matrix to the convex polyhedral cone consisting of all diagonally dominant matrices. It has a complexity of $O(p^2\log(p))$. See Section~\ref{sec:methodology}.} to project $[\bs{A}^{(t-1)} - \bs{J}^{(t-1)}]$ into the diagonally-dominant cone. Let $\bs{G}^{(t)}$ be the projected matrix.
\item Update $\bs{A}^{(t)}=\frac{1}{2}[\bs{G}^{(t-1)} + (\bs{G}^{(t-1)})^T]$ and $\bs{J}^{(t)} =\bs{J}^{(t-1)}+(\bs{G}^{(t)}-\bs{A}^{(t-1)})$.
\item If $\|\bs{J}^{(t)}-\bs{J}^{(t-1)}\|_F\leq \epsilon$, stop and output $\bs{A}=\bs{A}^{(t)}$.
\end{itemize}
\end{itemize}
This method is obtained by running one outer-loop iteration in the iterative algorithm to be introduced in Section~\ref{sec:methodology}, explaining the name of {\it one-step DD-PCA}. It has the same philosophy as the one-step Huber estimator \citep{bickel1975one} and one-step LLA implementation of non-convex penalized linear regressions \citep{zou2008one,fan2014strong}. It provides an approximate solution to \eqref{dd-PCA}, which is much faster to compute than solving \eqref{dd-PCA} exactly.

Exploring the approximate low-rank structures is a powerful strategy for big data analysis. The classical PCA has motivated many statistical methods. Similarly, DD-PCA and one-step DD-PCA can also serve as building blocks for statistical methodology development. We exemplify it in two statistical problems: the first is estimating a large covariance matrix, and the second is testing of the global null hypothesis in multiple testing.

Estimation of large covariance matrices is a popular topic in statistical literatures \citep{fan2016overview}. At the heart of it is two fundamental questions: (a) What structural assumption is appropriate? (b) How to evaluate the methods in real applications?

We adopt the structural assumption that the true covariance matrix $\bs{\Sigma}$ has an approximate low-rank decomposition with $\bs{A}\in \cSDD_c^+$. This is a special type of factor covariance structures that are commonly used in econometrics \citep{fan2008high}, finance \citep{fama1993common}, genetics  \citep{price2006principal} and many other fields. Our work is unique in the diagonal dominance assumption on $\bs{A}$. Intuitively, it is a natural relaxation of assuming $\bs{A}$ is diagonal, and it implies that, after the effects of latent factors are removed, the ``residual variables" are almost {\it uncorrelated}. Compared with existing covariance matrix estimators that assume $\bs{A}$ is sparse (e.g., \cite{fan2011high, poet}), this diagonal dominance structure benefits simultaneous estimation of $\bs{\Sigma}$ and $\bs{\Sigma}^{-1}$: In factor covariance structures, the singular values of the low-rank matrix are much larger than $\|\bs{A}\|$, so the error of estimating $\bs{\Sigma}$ is dominated by the error of recovering the low-rank part. If our goal is merely to estimate $\bs{\Sigma}$, we do not gain much from exploring the diagonal dominance structure of $\bs{A}$. However, if we are also interested in estimating $\bs{\Sigma}^{-1}$, the error of estimating $\bs{A}^{-1}$ will play a key role. Note that there always exists a matrix $\bs{B}\in\mathbb{R}^{n\times K}$ such that $\bs{L}=\bs{B}\bs{B}^{T}$. It follows from the matrix inverse formula \citep{horn2012matrix} that
\[
\bs{\Sigma}^{-1} = \bs{A}^{-1} - \bs{A}^{-1}\bs{B}(\bs{I}_K+\bs{B}^T\bs{A}^{-1}\bs{B})^{-1}\bs{B}^T\bs{A}^{-1}.
\]
Suppose we have obtained a good estimator $\widehat{\bs{\Sigma}}=\widehat{\bs{B}}\widehat{\bs{B}}^T + \widehat{\bs{A}}$ by fitting some factor covariance structure on the data. Even though $\|\widehat{\bs{\Sigma}}-\bs{\Sigma}\|$ is small, it is still possible that $\|\widehat{\bs{A}}^{-1}-\bs{A}^{-1}\|$ is large so that $\widehat{\bs{\Sigma}}^{-1}$ is far from being a good estimator of $\bs{\Sigma}^{-1}$. Fortunately, exploring the diagonal dominance structure largely mitigates this issue, thanks to an appealing feature of the diagonally-dominant cone $\cSDD_c^+$ \citep{horn2012matrix}:
\[
\|\bs{A}^{-1}\|\leq \frac{c}{c-1}\|[\mathrm{diag}(\bs{A})]^{-1}\|, \qquad\mbox{for any }\bs{A}\in \cSDD_c^+, \mbox{ where }c>1.
\]
Therefore, if we enforce $\widehat{\bs{A}}\in  \cSDD_c^+$ in fitting the factor covariance structure, for a constant $c>1$,
then $\|\widehat{\bs{A}}^{-1}\|$ won't explode, preventing ill behavior of $\widehat{\bs{\Sigma}}^{-1}$. To this end, we propose a new covariance matrix estimator $\widehat{\bs{\Sigma}}_{ddpca}$ using the solution of DD-PCA or one-step DD-PCA. We demonstrate in numerical studies:
$\widehat{\bs{\Sigma}}_{ddpca}$ has comparable performance with state-of-art methods, but the new estimator is tuning free once $K$ is specified, so is more convenient to use. Moreover, when it comes to estimating $\bs{\Sigma}^{-1}$ by $\widehat{\bs{\Sigma}}^{-1}_{ddpca}$, the new estimator is significantly better than inverting other factor-based covariance matrix estimators.

In real applications, estimating the covariance matrix is rarely the ultimate goal. Often, it serves as an intermediate step for downstream tasks. We demonstrate the usefulness of our covariance estimator by evaluating its performance in two downstream tasks, portfolio management and linear discriminant analysis. In the former, an estimate of the covariance matrix is needed to obtain Markowitz's optimal portfolio weights; in the latter, it is used to compute Fisher's LDA classifier. Note that what is actually plugged into these downstream tasks is the {\it inverse} of estimated covariance matrix. As we have argued, the main advantage of our method is on estimating $\bs{\Sigma}^{-1}$ by $\widehat{\bs{\Sigma}}_{ddpca}^{-1}$, a perfect match to these applications. This is supported by encouraging real data results.
It is worthwhile mentioning that our approach is different from the approach of plugging in an existing precision matrix estimator (e.g., the graphical lasso \citep{friedman2008sparse}). These methods assume $\bs{\Sigma}^{-1}$ is sparse, while we assume a factor structure on $\bs{\Sigma}$. For portfolio data, adopting a factor covariance structure is the common practice. For classification problems, there are also many real data sets for which it is appropriate to assume a factor structure (see Section~\ref{sec:CovEst}).

The second application of DD-PCA is for testing of the global null hypothesis in multiple testing. We are primarily interested in the setting where the mean vector in the alternative hypothesis is sparse. The Higher Criticism (HC) test  \citep{donoho2004higher}  is known to enjoy theoretical optimality and has gained increasing popularity in applications \citep{HCreviewDJ,wu2011rare}. However, the orthodox HC test assumes that the individual $z$-scores are mutually independent. There is limited understanding of how to extend the HC test to the case where $z$-scores share common latent factors. We model that the covariance matrix of $z$-scores, $\bs{\Sigma}$, has a low rank plus diagonal dominance decomposition. We propose two variants of HC for this setting, both utilizing DD-PCA as a module. The first test is a modification of the Innovated HC test \citep{hall2010innovated} by plugging in the estimator of $\bs{\Sigma}^{-1}$ from DD-PCA. The second test uses the low-rank matrix $\widehat{\bs{L}}$ from DD-PCA to remove the effects of latent factors and then applies the orthodox HC test. Both tests significantly improve over the orthodox HC test and the Innovated HC test in simulations. The rationale of these testing ideas is to ``transform" and ``decorrelate" the marginal $z$-scores. The first test extends the Innovated Transformation of $z$-scores \citep{hall2010innovated}, $X\mapsto \bs{\Sigma}^{-1}X$, from the case where $\bs{\Sigma}^{-1}$ is sparse to the case where $\bs{\Sigma}$ has a factor structure.
The second test aims to estimate and remove the latent factors from $z$-scores so that the `residuals' are almost uncorrelated (i.e., their covariance matrix is close to being diagonal). A similar idea was briefly mentioned in \cite{fan-han-gu-2012} (called
{\it factor-adjusted $z$-scores}), but it has never been used in the global testing problem.
In both new tests, we can replace DD-PCA by the classical PCA, but the numerical performance will deteriorate. This indicates that exploring the diagonal dominance structure is beneficial.

The remaining of this paper is organized as follows. In Section~\ref{sec:CovEst}, we introduce a new covariance matrix estimator powered by DD-PCA, and discuss its applications in portfolio optimization (Section~\ref{subsec:portfolio}) and linear discriminant analysis (Section~\ref{subsec:lda}). In Section~\ref{sec:Testing}, we propose two new test statistics, using DD-PCA as a building block, for testing of the global null hypothesis in multiple testing. In Section~\ref{sec:methodology}, we address the computation of DD-PCA, by introducing an ADMM algorithm and an iterative projection algorithm for conducting the decomposition \eqref{apprx-low-rank}-\eqref{assmp-A} for any given covariance matrix. Section~\ref{sec:simulation} contains simulations, and Section~\ref{sec:discussion} contains concluding remarks. The appendix contains some algorithm details omitted in Section~\ref{sec:methodology}.

\section{Estimating large covariance matrices by DD-PCA}\label{sec:CovEst}
Let $X\in\mathbb{R}^p$ be a multivariate random vector  with a covariance matrix $\bs{\Sigma}\in\mathbb{R}^{p\times p}$, where $p$ is presumably much larger than $n$. We adopt a factor model:
\be \label{factor-model}
X(j) = \sum_{k=1}^K b_{k}(j) W_k + Z(j), \qquad 1\leq j\leq p,
\ee
where $W_1,\ldots,W_K$ are unobserved random variables (factors), $b_k\in\mathbb{R}^p$ is a nonrandom vector containing the loadings of the $k$-th factor, and $Z\in\mathbb{R}^p$ is a random vector independent of the factors such that
\be \label{factor-model-A}
\bs{A}\equiv \mathrm{Cov}(Z) \in \cSDD^{+}.
\ee
Given $iid$ data $X_1,\ldots,X_n\in\mathbb{R}^p$, we are interested in estimating $\bs{\Sigma}$ and $\bs{\Sigma}^{-1}$.

By model \eqref{factor-model}-\eqref{factor-model-A}, the covariance matrix of $X$ has a decomposition
\[
\bs{\Sigma} = \bs{B}\mathrm{Cov}(W)\bs{B}^T + \bs{A}, \qquad \mbox{where}\quad \mathrm{rank}\bigl(\bs{B}\mathrm{Cov}(W)\bs{B}^T\bigr)=K \quad \mbox{and}\quad \bs{A}\in \cSDD^+.
\]
It has the low-rank plus diagonal dominance structure. We propose the following estimator: Let $\bs{S}=\frac{1}{n}\sum_{i=1}^n (X_i-\bar{X})(X_i-\bar{X})^T$ be the sample covariance matrix. Take $\bs{S}$ as the input to the one-step DD-PCA algorithm in Section~\ref{sec:intro} and let $(\widehat{\bs{L}},\widehat{\bs{A}})$ be the output. We estimate $\bs{\Sigma}$ by
\be \label{dd-CovEst}
\widehat{\bs{\Sigma}}_{ddpca}=\widehat{\bs{L}}+\widehat{\bs{A}}, \qquad \mbox{where $(\widehat{\bs{L}},\widehat{\bs{A}})$ is the output of one-step DD-PCA.}
\ee
We then estimate $\bs{\Sigma}^{-1}$ by the inverse of $\widehat{\bs{\Sigma}}_{ddpca}$.
Here, $(\widehat{\bs{L}},\widehat{\bs{A}})$ can be replaced by the output of other variants of DD-PCA (see Section~\ref{sec:methodology}). They give similar numerical performance, so we stick to one-step DD-PCA for computational convenience.

Different from existing covariance estimation methods under factor structures, our approach imposes the diagonal dominance constraint on $\bs{A}$. We now compare it with methods that impose the sparsity constraint on $\bs{A}$. One popular method is POET \citep{poet}. Let
$\bs{S}=\sum_{k=1}^p \lambda_k\xi_k \xi_k^T$
be the eigen-decomposition of $\bs{S}$, where $\lambda_k$ and $\xi_k$ are the $k$-th eigenvalue and eigenvector, respectively. The POET estimator is
\be  \label{poet}
\widehat{\bs{\Sigma}}_{poet}=\widehat{\bs{L}}_*+\mathcal{T}_a(\widehat{\bs{A}}_*),  \qquad \mbox{where}\quad \widehat{\bs{L}}_*=\sum_{k=1}^K\lambda_k \xi_k \xi_k^T, \quad \widehat{\bs{A}}_*=  \sum_{k=K+1}^p\lambda_k \xi_k \xi_k^T.
\ee
Here, $\mathcal{T}_a(\cdot)$ can be any entry-wise adaptive thresholding operator \citep{rothman2009,cai2011adaptive,xue2012positive}.
\cite{poet} suggest using the hard-thresholding operator applied to a ``correlation matrix" associated with $\widehat{\bs{A}}_*$, i.e.,
\be
\mathcal{T}_a(\widehat{\bs{A}}_*) = \widehat{\bs{D}}^{\frac{1}{2}} H_a\Bigl(\widehat{\bs{D}}^{-\frac{1}{2}} \widehat{\bs{A}}_* \widehat{\bs{D}}^{\frac{1}{2}}\Bigr) \widehat{\bs{D}}^{\frac{1}{2}} \qquad \text{where} \quad \widehat{\bs{D}} = \diag(\widehat{\bs{A}}_* ),
\ee
and $H_a(\cdot)$ is the entry-wise hard-thresholding at the threshold $a>0$.
Then, an estimate of $\bs{\Sigma}^{-1}$ is obtained by $\widehat{\bs{\Sigma}}^{-1}_{poet}$.

\begin{figure}[!t]
\centering
\includegraphics[width=\textwidth, trim=0 50 0 0, clip=true]{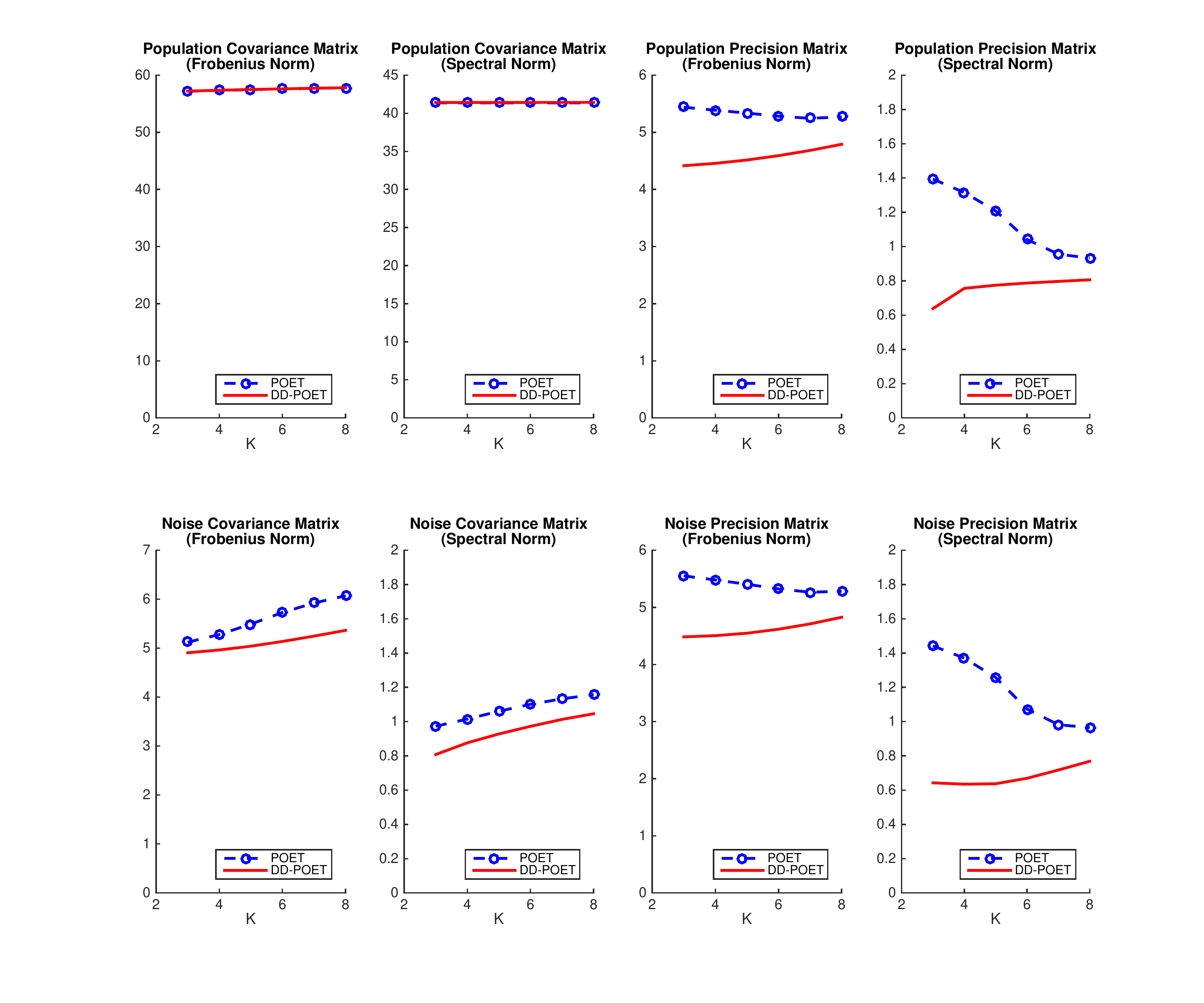} 
\caption{Comparison of our method with POET on estimating $\bs{\Sigma}$ (covariance matrix), $\bs{\Sigma}^{-1}$ (precision matrix), $\bs{A}$ (noise covariance matrix) and $\bs{A}^{-1}$ (noise precision matrix).}\label{fig:experiment}
\end{figure}

Figure~\ref{fig:experiment} gives a simulation example. Fix $(p,n,K)=(2000,200,3)$. We generate data from the model \eqref{factor-model}, where the factors $\{W_k(i): 1\leq k\leq K,1\leq i\leq n\}$ are drawn $iid$ from $N(0,1)$, the factor loadings $\{b_k(j): 1\leq k\leq K,1\leq j\leq p\}$ are generated $iid$ from ${\cal N}(0,1)$, and the noise vectors $Z_1,\ldots,Z_n$ are drawn $iid$ from a multivariate normal ${\cal N}_p(\bs{0}, \bs{A})$, with $A(i,j)=0.5^{|i-j|+1}$ for $i\neq j$ and $1$ otherwise. For both methods, $K$ is unknown and treated as a tuning integer. POET has an additional tuning threshold $a$, which is selected by cross-validation (default procedure in the {\it poet} package).\footnote{This default procedure guarantees that $\widehat{\bs{\Sigma}}_{poet}$ is invertible.}

In the top four panels of Figure~\ref{fig:experiment}, we show the average estimation errors on $\bs{\Sigma}$ and $\bs{\Sigma}^{-1}$ over $100$ repetitions. Since $K$ is unknown, we implement both methods for the true $K=3$ and misspecified $K\in \{4,5,\ldots,8\}$.\footnote{We don't include the results of $K\in \{1,2\}$, as the errors are much larger.}
For estimating $\bs{\Sigma}$, the two methods give very similar performance. This is not surprising. Since the eigenvalues of the low-rank part are much larger than $\|\bs{A}\|$, the error of estimating $\bs{\Sigma}$ is dominated by the error of recovering the low-rank part. Our method and POET has the same low-rank part (the $\widehat{\bs{L}}$ from one-step DD-PCA and the $\widehat{\bs{L}}_*$ in \eqref{poet} are indeed the same), so they have similar errors on estimating $\bs{\Sigma}$. From the bottom left two panels of Figure~\ref{fig:experiment}, we can see that our method does a better job on estimating $\bs{A}$; especially, the spectral norm error  is $10$-$20$\% smaller. However, this improvement is almost negligible compared with the errors on recovering the low-rank part. We conclude that our method and POET have similar performance on estimating $\bs{\Sigma}$. Still, our method has an advantage: It has no tuning threshold and is more convenient to use.

How about the performance on estimating $\bs{\Sigma}^{-1}$? The top right two panels of Figure~\ref{fig:experiment} clearly suggest that our method has a significant advantage. When $K=3$, the spectral norm error of our method is only one half of the error of POET. Interestingly, the performance of POET improves with an overshooting $K$; but even for $K=8$, its spectral norm error is still $10\%$ larger than the error of our method. For the Frobenius norm error, our estimator also outperforms POET for all choices of $K$. This phenomenon is due to that $\widehat{\bs{A}}$ plays a dominating role when we compute the inverse of $\widehat{\bs{\Sigma}}$, while the low-rank part has a negligible effect, so the advantage of our method on recovering $\bs{A}$ becomes prominent. This is illustrated in the bottom right two panels of Figure~\ref{fig:experiment}. Recall that $\widehat{\bs{A}}$ is from one-step DD-PCA and $\widehat{\bs{A}}_*$ is as in \eqref{poet}. The Frobenius/spectral norm of $(\widehat{\bs{A}}^{-1}-\bs{A}^{-1})$ is significantly smaller than the Frobenius/spectral norm of $(\widehat{\bs{A}}_*^{-1}-\bs{A}^{-1})$. Additionally, by comparing the top right two panels with the bottom right two panels, we can see that the error of estimating $\bs{\Sigma}^{-1}$ is almost determined by the error of estimating $\bs{A}^{-1}$.

This numerical example delivers two messages:  First, compared with competitive factor-based methods, the major advantage of our method is on estimating $\bs{\Sigma}^{-1}$ by $\widehat{\bs{\Sigma}}^{-1}$. Second, such an advantage is driven by the better accuracy on recovering $\bs{A}^{-1}$. Below, we explain them using linear algebra.

Without loss of generality, in model \eqref{factor-model}, we assume the covariance matrix of $W$ equals to the identify matrix. Then, $\bs{\Sigma}=\bs{B}\bs{B}^{T}+\bs{A}$. By matrix inverse formula,
\[
\bs{\Sigma}^{-1} = \bs{A}^{-1} - \bs{A}^{-1}\bs{B}(\bs{I}_K+\bs{B}^T\bs{A}^{-1}\bs{B})^{-1}\bs{B}^T\bs{A}^{-1}.
\]
Suppose we construct an estimator $\widehat{\bs{\Sigma}}=\widehat{\bs{B}}\widehat{\bs{B}}^T + \widehat{\bs{A}}$ from fitting a factor-type covariance structure. Then, $\widehat{\bs{\Sigma}}^{-1}$ (if it exists) has a similar decomposition:
\[
\widehat{\bs{\Sigma}}^{-1} = \widehat{\bs{A}}^{-1} - \widehat{\bs{A}}^{-1}\widehat{\bs{B}}(\bs{I}_K+\widehat{\bs{B}}^T\widehat{\bs{A}}^{-1}\widehat{\bs{B}})^{-1}\widehat{\bs{B}}^T\widehat{\bs{A}}^{-1}.
\]
By some basic linear algebra, we can derive the following proposition:
\begin{proposition}
Let $\widehat{\bs{A}}^{-\frac{1}{2}}\widehat{\bs{B}}=\sum_{k=1}^K\hat{\sigma}_k\hat{\eta}_k\hat{h}_k'$ be the singular value decomposition of $\widehat{\bs{A}}^{-\frac{1}{2}}\widehat{\bs{B}}$, where $\hat{\sigma}_k>0$ is the $k$-th singular value and $\hat{\eta}_k\in\mathbb{R}^p$ and $\hat{h}_k\in\mathbb{R}^K$ are the corresponding left and right singular vectors.
Then,
\be \label{inv-hatSigma}
\widehat{\bs{\Sigma}}^{-1} = \widehat{\bs{A}}^{-1} - \widehat{\bs{A}}^{-\frac{1}{2}}\biggl(\sum_{k=1}^K \frac{1}{\hat{\sigma}_k^{-2}+1}\hat{\eta}_k\hat{\eta}_k' \biggr) \widehat{\bs{A}}^{-\frac{1}{2}}
\ee
\end{proposition}

By \eqref{inv-hatSigma}, the error of recovering the low-rank part only affects the matrix in the brackets. For $(\widehat{\bs{A}},\widehat{\bs{B}})$ obtained in factor-based methods, nonzero eigenvalues of $\widehat{\bs{B}}\widehat{\bs{B}}^T$ are much larger than $\|\widehat{\bs{A}}\|$, so $\widehat{\sigma}_k$'s are all very large.
Then, the matrix in the brackets can hardly bring in a large error in $\widehat{\bs{\Sigma}}^{-1}$. The error in
$\widehat{\bs{\Sigma}}^{-1}$ mainly comes from the error in $\widehat{\bs{A}}^{-1}$.

We further investigate the error in $\widehat{\bs{A}}^{-1}$. Note that
\be
\| \widehat{\bs{A}}^{-1} - \bs{A}^{-1}\|\leq \|\widehat{\bs{A}}^{-1}\|\|\bs{A}^{-1}\|\| \widehat{\bs{A}} - \bs{A} \|.
\ee
To achieve a small $\| \widehat{\bs{A}} - \bs{A} \|$ by imposing structural assumptions on $\bs{A}$ is not too difficult. However, it typically does not prevent $\|\widehat{\bs{A}}^{-1}\|$ from exploding. For example, if $\widehat{\bs{A}}$ is obtained from entry-wise thresholding, we need a comparably large threshold to control $\|\widehat{\bs{A}}^{-1}\|$, but unfortunately we cannot let the threshold be too large as it significantly increases $\| \widehat{\bs{A}} - \bs{A} \|$.
It turns out that, if we restrict $\widehat{\bs{A}}\in \cSDD_c^+$ for a constant $c>1$, then it is automatically guaranteed that $\|\widehat{\bs{A}}^{-1}\|$ has a nice bound. As a property of diagonally-dominant matrices \citep{horn2012matrix}, for $c>1$,
\[
\lambda_{\min}(\widehat{\bs{A}}) \geq \min_{1\leq j\leq p}\Bigl\{ \widehat{a}_{jj}- \sum_{i: i\neq j}|\widehat{a}_{ji}|\Bigr\}
\geq \min_{1\leq j\leq p}\Bigl\{ \widehat{a}_{jj} - c^{-1}\widehat{a}_{jj}\Bigr\}\geq \frac{c-1}{c}\min_{1\leq j\leq p}\widehat{a}_{jj}.
\]
It follows that
\be \label{SDDprop}
\|\widehat{\bs{A}}^{-1}\|\leq \frac{c}{c-1}\| [\mathrm{diag}(\widehat{\bs{A}})]^{-1} \|.
\ee
This explains why the constraint of $\widehat{\bs{A}}\in \cSDD_c^+$ helps significantly reduce the errors in $\widehat{\bs{A}}^{-1}$ and (ultimately) the errors in $\widehat{\bs{\Sigma}}^{-1}$.

The above argument applies to $c>1$. In our method, $c=1$. Sometimes, we may even have to use $c<1$, so that the assumption $\bs{A}\in \cSDD_c^+$ is not too restrictive (see Section~\ref{sec:methodology}). For $c\leq 1$, we do not have a solid argument as \eqref{SDDprop}, but a similar phenomenon is observed in numerical studies.

Below, we use two real applications to further demonstrate that exploring the diagonal dominance factor structures is a useful strategy.

\subsection{Application to portfolio management} \label{subsec:portfolio}
Given a collection of $p$ assets, portfolio management aims to determine the weights allocated to each asset. It is often desirable to construct the {\it minimum risk portfolio}, where the asset weights $\bs{w}^*=(w_1^*,\ldots,w_p^*)$ are determined by
\[
\bs{w}_* = \mathrm{argmin}_{\bs{w}^T\bs{1}=1}  \bs{w}^T\bs{\Sigma}\bs{w}, \qquad  \mbox{$\bs{\Sigma}\in\mathbb{R}^{p\times p}$: asset covariance matrix}.
\]
In practice, $\bs{\Sigma}$ is unknown. We first obtain an estimate $\widehat{\bs{\Sigma}}$ using asset returns $\bs{y}_1,\ldots,\bs{y}_n\in\mathbb{R}^p$ during a period of $n$ days, then we estimate the weights by
\[
\widehat{\bs{w}}_* = \mathrm{argmin}_{\bs{w}^T\bs{1}=1}  \bs{w}^T\widehat{\bs{\Sigma}}\bs{w}.
\]
This optimization has an explicit solution:
\be \label{portfolio}
\widehat{\bs{w}}_* = (\bs{1}^T\widehat{\bs{\Sigma}}^{-1}\bs{1})^{-1}(\widehat{\bs{\Sigma}}^{-1}\bs{1}).
\ee
Since what we actually need is $\widehat{\bs{\Sigma}}^{-1}$, exploring the low-rank plus diagonal dominance structure is a potentially useful strategy.

\begin{figure}[!t]
\centering
\includegraphics[width=0.7\textwidth, trim=0 20 0 0, clip=true]{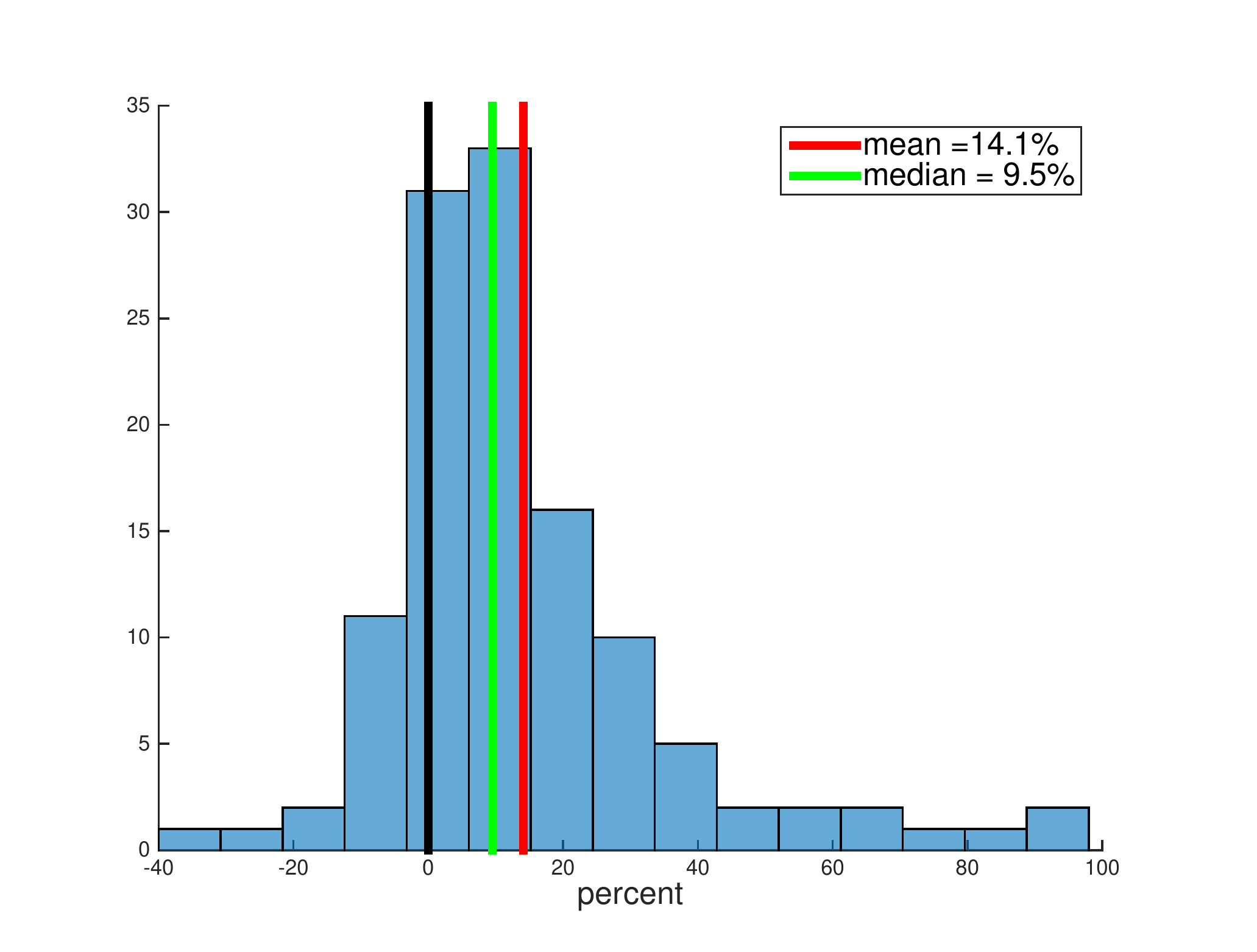} \vspace{-15pt}
\caption{Histogram of ratio of improvement of our method over POET over $120$ months.}\label{fig:portfolio}
\end{figure}

We compare our method with POET on real data. We collected the daily returns of stocks in S\&P 100 index from January 1st 2006 to December 31st 2016.
After removing companies that were listed after 2006, there are $80$ stocks in total. On the first trading day of each month, we created two portfolios from \eqref{portfolio}, where $\widehat{\bs{\Sigma}}$ is estimated using daily returns for the proceeding 12 months ($n=252$) by our method and by POET, respectively. We set $K=3$ for both methods. The threshold in POET is chose by cross-validation (we use the default cross-validation procedure in {\it poet} package). On the last trading day of the same month, we measure the actual risk of each portfolio by
\[
R(\widehat{\bs{w}}_*) = \frac{1}{T}\sum_{t=1}^T (\bs{y}_t^T \widehat{\bs{w}}^*)^2,
\]
where $T$ is the number of trading days in this month ($T=21$ for most months) and $\bs{y}_t\in\mathbb{R}^{80}$ contains the stock returns on day $t$ of the month.
Define $r=(R_{poet}-R_{ddpca})/R_{ddpca}$; note that a positive $r$ indicates that the portfolio created using our method is superior to that of POET. Figure~\ref{fig:portfolio} displays the histogram of $r$ over $120$ months in our data range.
It suggests that our method improves POET by $9.5\%$ on average and $14.7\%$ in the median.

\subsection{Application to linear discriminant analysis} \label{subsec:lda}
In binary classification, given feature vectors $X_1,\ldots,X_n\in\mathbb{R}^p$ and training labels $\ell_1,\ldots,\ell_n\in \{1,2\}$, we aim to construct a linear classifier. In the classical regime where $p$ is fixed as the training sample size grows, Fisher's LDA is an effective linear classifier. In the modern high dimensional settings where $p\gg n$, it has been well understood that feature screening is necessary before one applies Fisher's LDA \citep{fan2008high,donoho2008higher}, and that it is desirable to plug in a good estimate of the inverse covariance matrix that explores structural assumptions \citep{cai-liu-luo-2011}. Recently, \cite{fan2013optimal} proposed a linear classifier that uses an estimate of inverse covariance matrix in both the screening step and LDA step, and they showed that this classifier is rate-optimal under a multivariate normal model with even extremely weak signal strength. This classifier was later applied to several large real classification problems with superior results \citep{huang2016partial}. We shall combine our covariance matrix estimator with this classifier to see whether exploring the low-rank plus diagonal-dominance structure is helpful.

Given an estimate $\widehat{\bs{\Omega}}$ of the inverse covariance matrix and a threshold $t>0$, the classifier has four steps \citep{huang2016partial}:
\begin{enumerate}
\item Calculate the feature-wise $t$-score: For $1\leq j\leq p$, let $Z(j) = [\bar{X}_1(j) - \bar{X}_2(j)]/(n \cdot s_j) $, where $\bar{X}_1(j)$ and $\bar{X}_2(j)$ are the within-class sample means of feature $j$ and $s_j>0$ is the pooled standard deviation of feature $j$. Write $Z=(Z(1),\ldots,Z(p))^T$.
\item Apply the Innovated Transformation \citep{fan2013optimal} to get $\tilde{Z}=\widehat{\bs{\Omega}}Z$.
\item Feature-wise thresholding: For $1\leq j\leq p$, let $w(j) = \text{sgn}(\tilde{Z}(j) ) \cdot 1\{|\tilde{Z}(j) | \geq t \}$. Write $w = (w(1), w(2), \dots, w(p))^{T}$.
\item Classification by LDA. Given a test feature vector $\widetilde{X}\in\mathbb{R}^p$, for $1\leq j\leq p$, normalize $\widetilde{X}(j)$ to $\widetilde{X}^{\ast}(j)= [\tilde{X}(j) - \frac{1}{2}(\bar{X}_1(j) + \bar{X}_2(j) )]/s_j$, where $(\bar{X}_1(j), \bar{X}_2(j), s_j)$ are the same as in Step 1. Write $\widetilde{X}^{\ast}=(\widetilde{X}^{\ast}(1),\ldots,\widetilde{X}^{\ast}(p))^T$. We classify the test sample to class 1 if $ w^T \widehat{\bs{\Omega}} \widetilde{X}^{\ast}>0$ and to class 2 otherwise.
\end{enumerate}

In this classifier, the matrix $\widehat{\bs{\Omega}}$ plays two roles: First, it is used in the Innovated Transformation, so different $\widehat{\bs{\Omega}}$ leads to different feature rankings. Second, it is used in the LDA step, so $\widehat{\bs{\Omega}}$ also affects the classification boundary.

We compare the classification performance of plugging in three versions of $\widehat{\bs{\Omega}}$: The first is $\widehat{\bs{\Sigma}}_{ddpca}^{-1}$, the second is $\widehat{\bs{\Sigma}}^{-1}_{poet}$, and the last is $[\mathrm{diag}(\bs{S})]^{-1}$, where $\bs{S}$ is the sample covariance matrix. We note that the last approach is indeed the method FAIR \citep{fan2008high}. The above classifier also requires a threshold $t>0$. To minimize the effects of selecting $t$, for each $1\leq k\leq p$, we set the threshold such that $k$ features are retained and record the classification error. This generates an error curve for each method as $k$ ranges from $1$ to $p$.

We consider two datasets: the lung cancer dataset \citep{gordon2002translation} and the breast cancer dataset \citep{wang2005gene}. They were downloaded from \url{http://blog.nus.edu.sg/staww/softwarecode/}. For both datasets, we  conducted a pre-processing by ranking all features by the feature-wise $t$-score and retaining $p_0$ top-ranked features, where $p_0$ is a number that is for sure larger than the true number of useful features (but $p_0\ll p$).
\begin{center}
\begin{tabular}{ccccc}
\hline
dataset & sample size & dimension & $p_0$\\
\hline
Lung cancer & 181 & 12,533 & 100\\
breast cancer & 276 & 22,215 & 1000\\
\hline
\end{tabular}
\end{center}
The lung cancer dataset was analyzed in various papers \citep{tibshirani2002diagnosis,fan2008high}. The estimated number of useful features by these methods is around $30$, so we confidently set $p_0=100$. The breast cancer dataset is a more difficult one and requires a lot more retained features. \cite{jin2016influential} analyzed the dataset under a clustering framework and suggested that the number of useful features is 728, so we set $p_0=1000$.
We also tried other choices of $p_0$ (e.g., $p_0=200$ for lung cancer data and $p_0=2000$ for breast cancer data), and the results are similar.

We evaluate the classification performance by a 5-fold cross-validation procedure with stratified sampling. In detail, we randomly divide samples from class 1 into five folds and do the same to samples from class 2; we then re-combine them to five folds, such that the fraction of class 1 is the same across all folds. Next, we successively leave out each fold, train the classifier on remaining samples, and compute the test error on leave-out samples. The misclassification error reported is the average over 5 folds.

\begin{figure}[!t]
\centering
\includegraphics[width=0.9\textwidth]{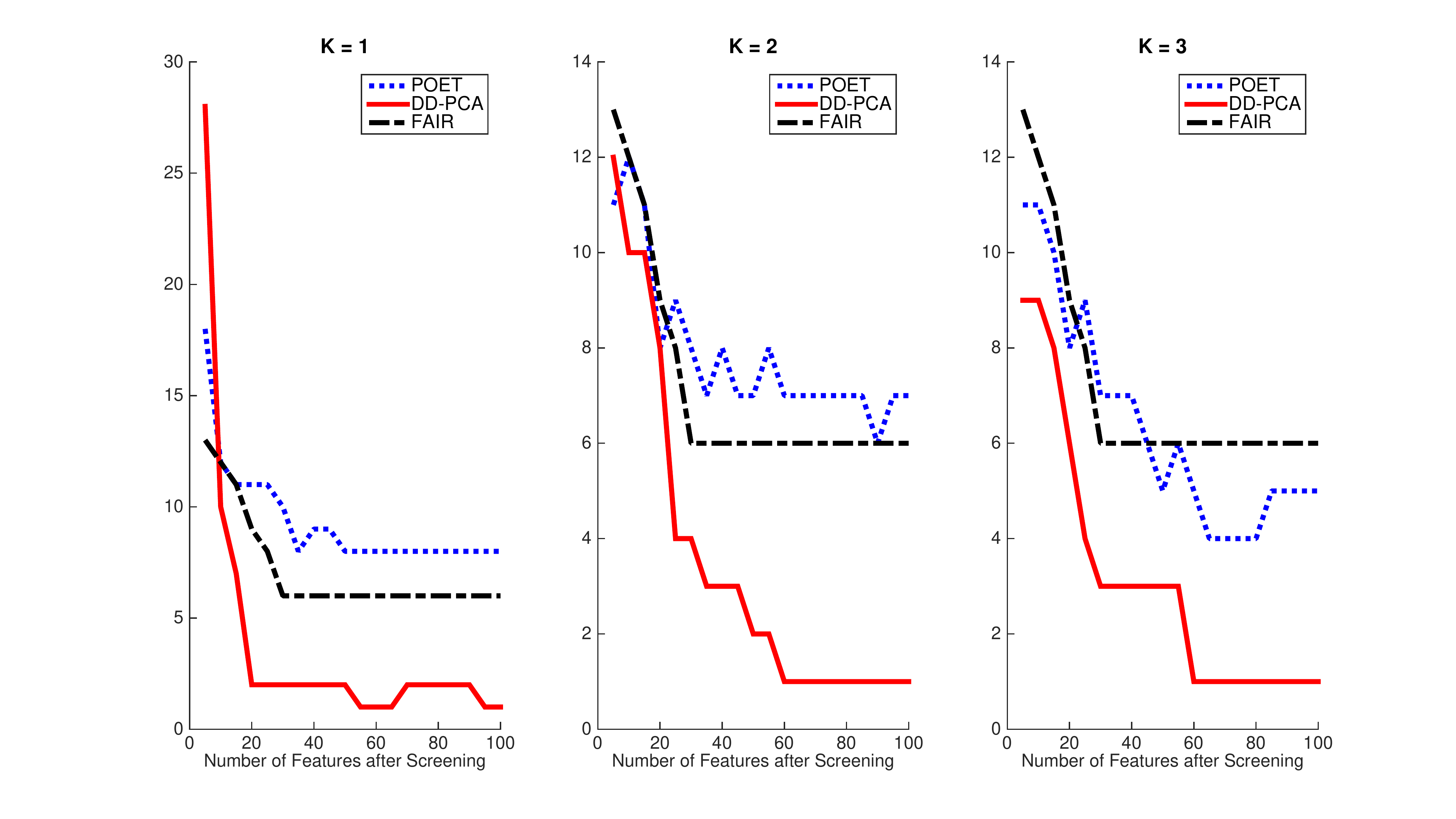} \vspace{-20pt}
\caption{Misclassification errors on lung cancer data ($n=181$).} \label{fig:LDA_lung}
\end{figure}

Figure~\ref{fig:LDA_lung} displays the results on lung cancer dataset. POET and DD-PCA have a tuning integer $K$, and we tried $K\in \{1,2,3\}$. The results suggest that, as long as more than $10$ features are retained, the classifier powered by DD-PCA uniformly outperforms the other two. Especially, for $K\in \{2,3\}$, the error keeps as low as
$1/181$ once the number of retained features exceeds $60$. The performance of POET is slightly worse than FAIR for $K\in \{1,2\}$, and slightly better for $K=3$. We emphasize that the estimated inverse covariance matrix affects both the feature ranking and the LDA; therefore, even when the number of retained features is the same, the actual retained features are different across different methods.

\begin{figure}[!t]
\centering
\includegraphics[width=0.9\textwidth]{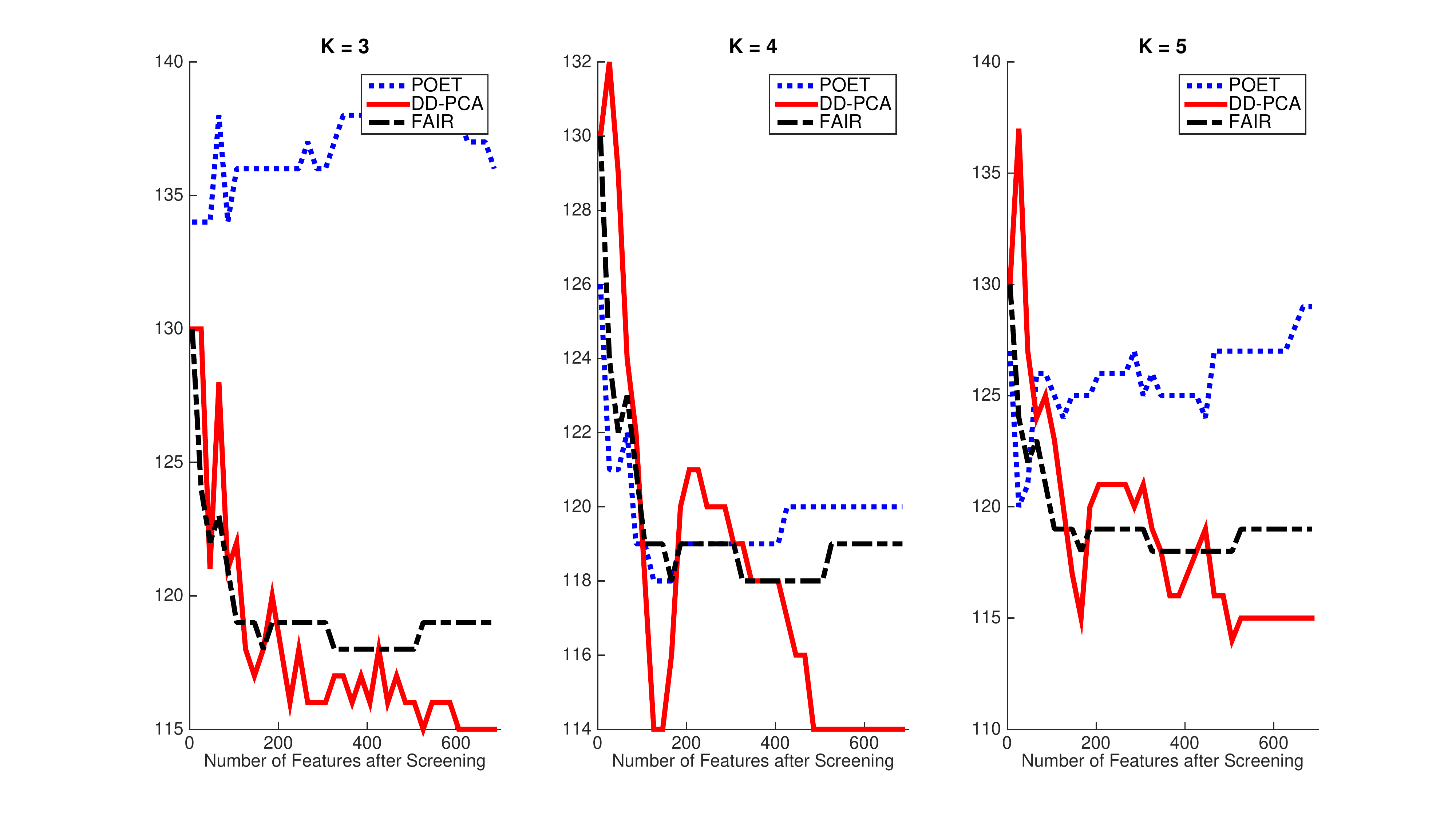} \vspace{-20pt}
\caption{Results for breast cancer data $(n= 276)$} \label{fig:LDA_breast}
\end{figure}

Figure~\ref{fig:LDA_breast} displays the results on breast cancer dataset. For both POET and DD-PCA, $K\in \{4,5\}$ is favored to $K=3$. When $K=4$, as the number of retained features is in the interval of $[500,700]$, DD-PCA achieves the smallest error of $114/276$. In all three panels, the lowest attainable error of DD-PCA is smaller than those of POET and FAIR.

\section{Detecting sparse mean effects by DD-PCA}  \label{sec:Testing}
The global detection is a problem of great interest in multiple testing \citep{simes1986improved,donoho2004higher,wu2011rare}. Let $X_1,\ldots,X_p\in\mathbb{R}$ be the $z$-scores of $p$ tests, where $p$ is presumably large. We assume
\be \label{testing-mod}
X \sim \cN_p(\mu, \bs{\Sigma}),
\ee
where $\mu$ contains the true effects of these tests and $\bs{\Sigma}$ captures the dependence among the $z$-scores.
We are interested in testing
\be \label{testing-hypothesis}
H_0: \mu=0, \qquad v.s. \qquad H_1: \mbox{$\mu\neq 0$, and $\mu$ is sparse}.
\ee

When $\bs{\Sigma}$ is a diagonal matrix, this problem has been studied extensively in the literature. Various tests were proposed, such as the $\chi^2$ test (or Hotelling's $T^2$ test), maximum entry test (or minimum $p$-value test), Higher-Criticism test \citep{donoho2004higher}, Berk-Jones test \citep{jager2007goodness}, etc.. The Higher-Criticism (HC) test achieves the theoretically optimal detection boundary when the nonzero effects in $\mu$ are rare and weak \citep{donoho2004higher} and has gained increasing popularity in real applications \citep{HCreviewDJ}. However, there is limited understanding of how to use the HC test beyond a diagonal $\bs{\Sigma}$. When $\bs{\Sigma}$ is heavily non-sparse, a brute-forth application of the orthodox HC test leads to suboptimal performance \citep{HCdepend}. A satisfactory answer is only available when $\bs{\Sigma}$ is row-wise sparse. \cite{hall2010innovated} introduced the ``Innovated Transformation" on data, $X \mapsto \widehat{\bs{\Omega}}X$, where $\widehat{\bs{\Omega}}$ is an estimator of $\bs{\Sigma}^{-1}$. They showed that an application of the HC test on the post-transformation data lead to theoretically optimal testing performance.

Motivated by the popularity of adopting factor covariance structures in multiple testing \citep{leek-storey-2008,fan-han-gu-2012}, we consider the global testing problem \eqref{testing-mod}-\eqref{testing-hypothesis} by assuming that $\bs{\Sigma}$ has a low rank plus diagonal dominance structure as in \eqref{apprx-low-rank}-\eqref{assmp-A}. Denote by $\sum_{k=1}^K \nu_k \eta_k\eta_k^T$ the eigen-decomposition of $\bs{L}$. Equivalently, we model that
\begin{equation} \label{TestModel}
X = \mu + \sum_{k=1}^K w_k \eta_k + z, \quad w_k \sim \mathcal{N}(0,\nu_k),\; z \sim \mathcal{N}_p (0, \bs{A}),\; \mbox{$w_1,\ldots,w_K,z$ are independent}.
\end{equation}
Here, $w_1,\ldots,w_K$ are latent variables that account for most of the heavy dependence among test statistics. Under this model, $\bs{\Sigma}$ is heavily non-sparse, so the orthodox HC test performs unsatisfactorily \citep{HCdepend}. At the same time, $\bs{\Sigma}^{-1}$ may not be row-wise sparse, so the Innovated HC test \citep{hall2010innovated} is not necessarily a good choice either. How to adapt the HC test to factor covariance structures is still largely unclear. We use DD-PCA (and its variants such as one-step DD-PCA) to develop two modifications of the HC test. Both tests significantly outperform the orthodox HC and Innovated HC, when the factor covariance structure holds.

Without loss of generality, we assume an estimate of $\bs{\Sigma}$ is available, denoted as $\widehat{\bs{\Sigma}}$. Note that it is common that a $z$-score $X_j$ is computed from a number of repeated observations. Suppose we observe $iid$ samples $X_1,X_2,\ldots,X_n$ and obtain the $z$-scores as $X=\frac{1}{\sqrt{n}}\sum_{i=1}^n X_i$, then the sample covariance matrix of $X_i$'s can be used as $\widehat{\bs{\Sigma}}$.

We begin with describing the orthodox HC. Let $\pi_1,\pi_2,\ldots,\pi_p$ be the marginal $P$-values computed from $X_j\sim {\cal N}(0,\widehat{\Sigma}_{jj})$, where $\widehat{\Sigma}_{jj}$ is the $j$-th diagonal of $\widehat{\bs{\Sigma}}$. Sort the $P$-values in the descending order and denote by $\pi_{(k)}$ the $k$-th smallest $P$-value, $1\leq k\leq p$. The OHC test statistic is
\begin{equation} \label{OHC}
HC_{p}^*=\max_{1\leq j\leq p/2}HC_{p,j},\qquad \mbox{where}\quad HC_{p,j} = \frac{\sqrt{p}[(j/p)-\pi_{(j)}]}{\sqrt{\pi_{(j)}(1-\pi_{(j)})}}, \qquad 1\leq j\leq p.
\end{equation}
The null distribution of $HC_p^*$ is often approximated by a Gumbel distribution or by simulating the null data \citep{HCreviewDJ}. Although OHC was proposed for the case of a diagonal $\bs{\Sigma}$, we can treat it as a blackbox procedure (we actually know what is happening in the `blackbox', but we have no intention to interfere): It takes as input the  $P$-value for each individual test, and outputs a test statistic for the global null hypothesis which is an aggregation of all individual $P$-values. Now, if we feed this `blackbox' with a different set of individual $P$-values, it will output a different test statistic. Following this strategy, we modify OHC by constructing individual $P$-values using $\widehat{\bs{\Sigma}}$, in hopes of borrowing strength from each other.

\begin{figure}[H]
\centering
\includegraphics[width=1.05\textwidth]{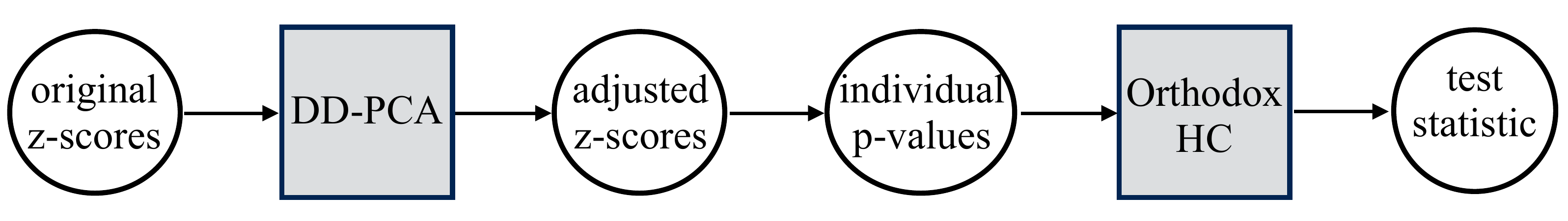}
\caption{Illustration of the use of DD-PCA to modify the orthodox HC.}
\end{figure}

The first test we propose is {\it IHC-DD test}, where `IHC' stands for Innovated HC and `DD' stands for DD-PCA. As the name has suggested, this test is built on top of the Innovated HC test---we replace $\widehat{\bs{\Omega}}$ in IHC by the DD-PCA estimator of $\bs{\Sigma}^{-1}$ introduced in Section~\ref{sec:CovEst}.
\begin{itemize}
\item Take $\widehat{\bs{\Sigma}}$ as the input to the one-step DD-PCA algorithm in Section~\ref{sec:intro} and let $(\widehat{\bs{L}},\widehat{\bs{A}})$ be the output. Let $\widehat{\bs{\Sigma}}_{ddpca}=\widehat{\bs{L}}+\widehat{\bs{A}}$.
\item Obtain $\widetilde{X}=\widehat{\bs{\Sigma}}_{ddpca}^{-1}X$. Compute the indivisual $P$-values $\widetilde{\pi}_j$ from the null distribution of $\widetilde{X}_j\sim {\cal N}(0, \widehat{\Omega}_{jj})$, where $\widehat{\Omega}_{jj}$ is the $j$-th diagonal of $\widehat{\bs{\Sigma}}_{ddpca}^{-1}$.
\item Input the $P$-values $\widetilde{\pi}_1,\ldots,\widetilde{\pi}_p$ to the OHC procedure \eqref{OHC} to get a test statistic. \end{itemize}
The individual $P$-values fed to OHC are different from before: Each $\widetilde{\pi}_j$ borrows information from $z$-scores of other tests, taking advantage of the dependence between tests. The IHC-DD test can be viewed as another application of the DD-PCA covariance estimator proposed in Section~\ref{sec:CovEst}, where we simply plug the estimated $\bs{\Sigma}^{-1}$ into the existing IHC test.

The second test we propose is more customized to the factor covariance structure. We call it {\it DD-HC} test, to  differentiate it from the test above. By \eqref{TestModel},
\[
X - \sum_{k=1}^K w_k \eta_k \sim {\cal N}_p(\mu, \bs{A}).
\]
Provided with estimates of $w_k$'s, $\eta_k$'s, and $\bs{A}$, we can use $X^*=X-\sum_{k=1}^K\widehat{w}_k\widehat{\eta}_k$ as the new $z$-scores for all individual tests, and we can compute the individual $P$-values from the null distribution of $X_j^*\sim {\cal N}(0, \widehat{A}_{jj})$, where $\widehat{A}_{jj}$ is the $j$-th diagonal of $\widehat{\bs{A}}$. Let's discuss how to estimate $(w_1,\ldots,w_K,\eta_1,\ldots,\eta_K, \bs{A})$ from DD-PCA. Let $(\widehat{\bs{L}}, \widehat{\bs{A}})$ be the output of DD-PCA with $\widehat{\bs{\Sigma}}$ as the input (we will discuss below which DD-PCA algorithm to use). Recall that in model \eqref{TestModel}, $\eta_k$'s are eigenvectors of $\bs{L}$. This motivates us to estimate $\eta_k$'s by the leading eigenvectors of $\widehat{\bs{L}}$. Once we have $\widehat{\eta}_1,\ldots,\widehat{\eta}_K$, we can approximate model \eqref{TestModel} by
\[
X = \mu+ \sum_{k=1}^K w_k\widehat{\eta}_k + z.
\]
This is indeed a linear regression model with $K$ covariates and $p$ observations. The regression coefficients are $w_1,\ldots,w_K$, and each observation has an individual intercept $\mu_j$. Since $\mu$ is a sparse mean vector, the majority of $\mu_j$'s are zero; we thus estimate $w_1,\ldots,w_K$ by a robust regression via minimizing $\|X-\sum_{k=1}^K w_k\widehat{\eta}_k\|_1$, with respect to $w_k$'s. Last, we discuss the estimation of $\bs{A}$. A straightforward idea is to use $\widehat{\bs{A}}$ output by DD-PCA. However, we don't recommend this approach. The estimator of $\bs{A}$ shall be used to approximate the null distribution of $X_j^*=X_j-\sum_{k=1}^K\widehat{w}_k\widehat{\eta}_k(j)$, in order to compute individual $P$-values. So, our goal is not to estimate $\bs{A}$ accurately but to provide an accurate estimate of the variance of $X_j^*$. We have to take into account the plug-in effect of $\widehat{w}_k$ and $\widehat{\eta}_k$. Since $X_j^*$ is the residual of fitting a linear regression, a more reasonable choice is to use the diagonals of $\widehat{\bs{\Sigma}}-\widehat{\bs{L}}$ as estimates of the variances of $X_j^*$'s. We summarize the procedure as follows:
\begin{itemize}
\item Take $\widehat{\bs{\Sigma}}$ as the input to a DD-PCA algorithm and and let $(\widehat{\bs{L}},\widehat{\bs{A}})$ be the output.
\item Conduct PCA on $\widehat{\bs{L}}$, and denote by $\widehat{\eta}_1,\ldots,\widehat{\eta}_K$ the first $K$ eigenvectors of $\widehat{\bs{L}}$.
\item Regress $X$ on $(\widehat{\eta}_1,\ldots,\widehat{\eta}_K)$ using a robust regression: $\min_{w_1,\ldots,w_K}\|X - \sum_{k=1}^K w_k\widehat{\eta}_k\|_1$. Let $(\widehat{w}_1,\ldots,\widehat{w}_K)$ be the estimated coefficients.
\item Obtain $X^*=X-\sum_{k=1}^K\widehat{w}_k\widehat{\eta}_k$. Compute the individual $P$-values $\pi^*_j$ from the null distribution of $X_j^*\sim {\cal N}(0, \widehat{R}_{jj})$, where $\widehat{R}_{jj}$ is the $j$-th diagonal of $\widehat{\bs{R}}\equiv\widehat{\bs{\Sigma}}-\widehat{\bs{L}}$.
\item Input the $P$-values $\pi^*_1,\ldots,\pi^*_p$ to the OHC procedure \eqref{OHC} to get a test statistic.
\end{itemize}
Since the variance of $X_j^*$ is smaller than the variance of $X_j$, we expect that feeding to OHC with these new $P$-values $\pi_j^*$ will lead to more testing power. Now, let's consider the choice of the DD-PCA algorithm. We shall use the iterative projection algorithm to be introduced in Section~\ref{subsec:iter-alg}. The one-step DD-PCA in Section~\ref{sec:intro} is equivalent to running this iterative algorithm with only one iteration. For the sake of constructing the DD-HC test statistic, we need more iterations. By design of the iterative algorithm, as the number of iterations increases, $\|\widehat{\bs{\Sigma}}-\widehat{\bs{L}}-\widehat{\bs{A}}\|_F$ continues to decrease.  It indicates that $\widehat{\bs{R}}$ and $\widehat{\bs{A}}$ are closer to each other, and so $\widehat{\bs{R}}$ becomes more diagonal dominant (note that $\widehat{\bs{A}}$ is forced to be in the diagonal dominant cone). The matrix $\widehat{\bs{R}}$ approximately captures the dependence structure in $X^*$. When $\widehat{\bs{R}}$ is close to being diagonal, it means the dependence among original $z$-scores has been fully utilized and there is minimal loss by ignoring the dependence among entries of $X_j^*$.

{\bf Remark}. We are not the first to consider using a factor covariance structure to adjust $P$-values in multiple testing. \cite{fan-han-gu-2012} proposed a similar idea, but their $(\widehat{\bs{L}}, \widehat{\bs{A}})$ are from the classical PCA. Our innovation is two fold. First, we are the first to incorporate a factor covariance structure in testing against the global null hypothesis. The main focus of \cite{fan-han-gu-2012} is on estimating the false discovery proportion (FDP). Although they briefly mentioned the idea of computing factor-adjusted $P$-values, they didn't use it in their method (their FDP estimator is still based on the original $P$-values). Second, our method is equipped with the fresh DD-PCA algorithm, in contrast with the classical PCA used in \cite{fan-han-gu-2012}.

\begin{figure}[t]
\centering
\includegraphics[width=.495\textwidth, trim=30 0 30 0, clip=true]{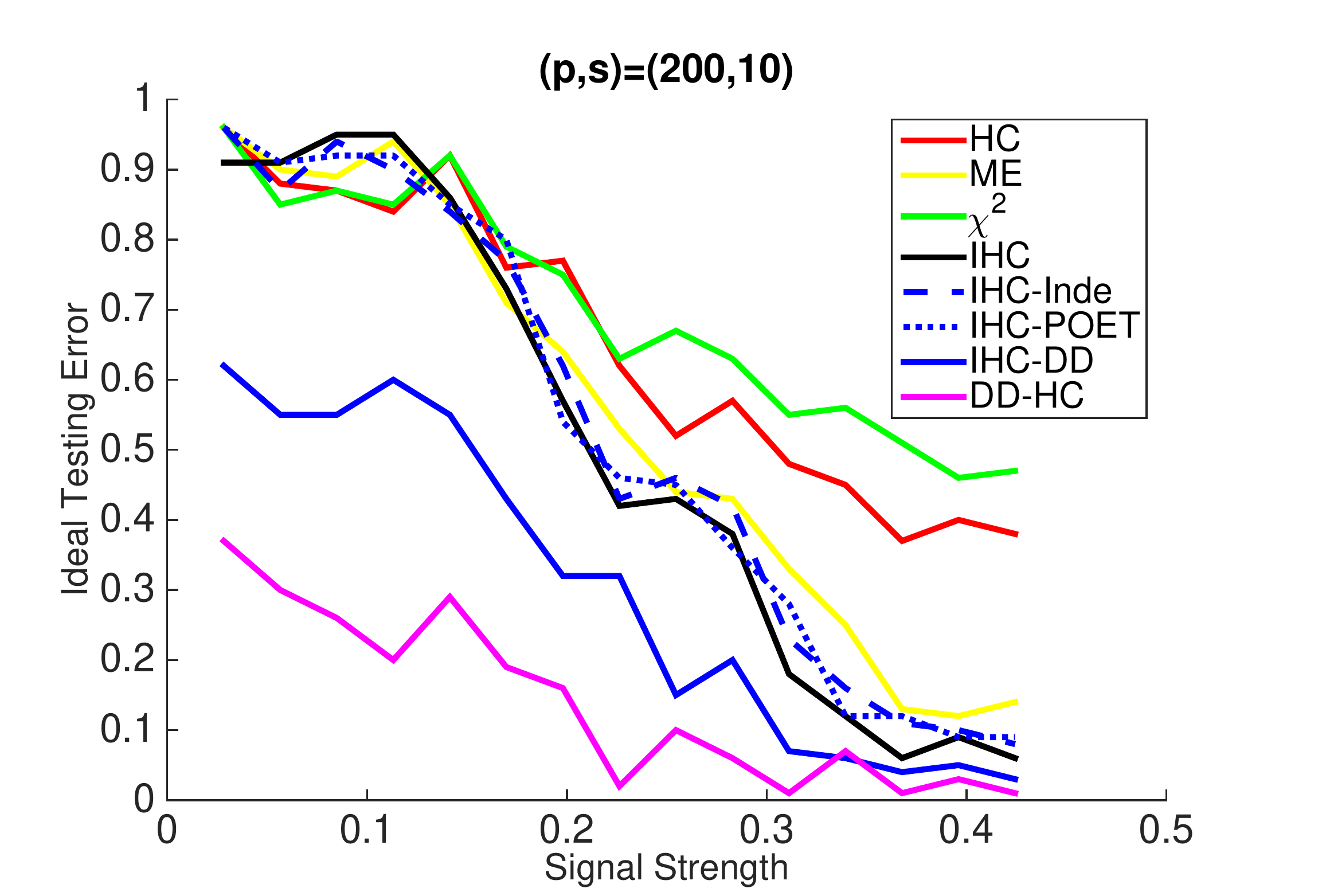}
\includegraphics[width=.495\textwidth, trim=30 0 30 0, clip=true]{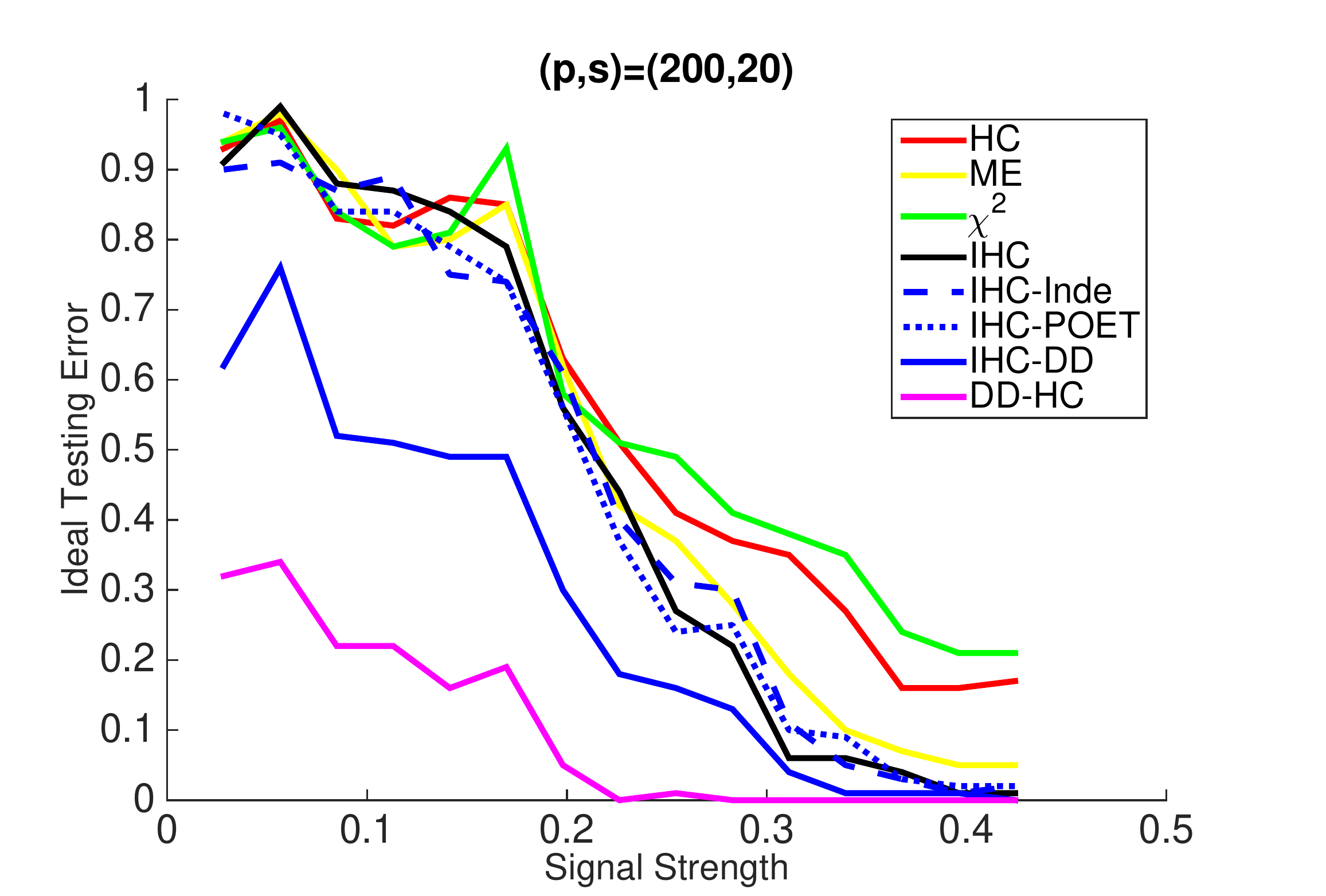}\\
\vspace{5pt}
\includegraphics[width=.495\textwidth, trim=30 0 30 0, clip=true]{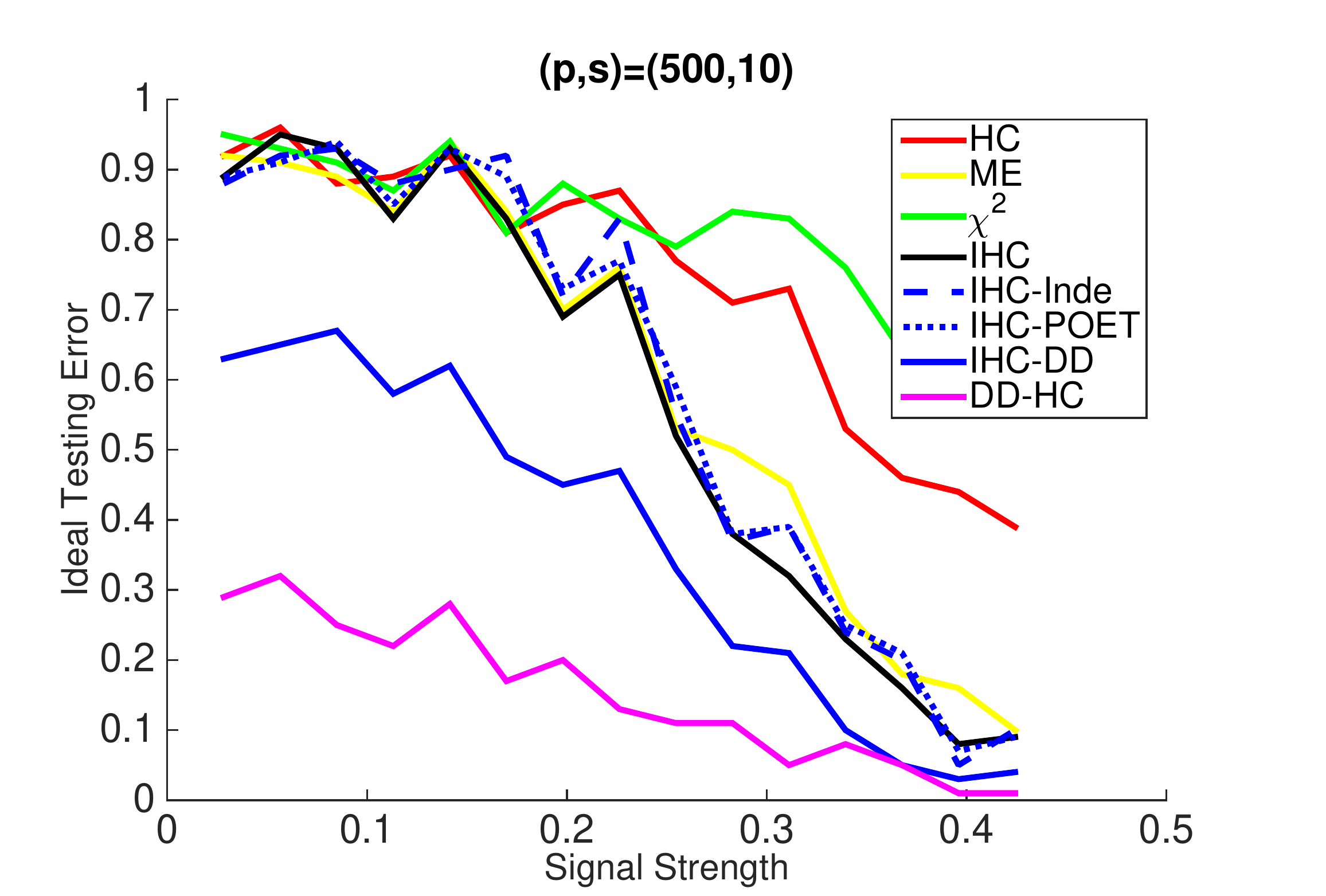}
\includegraphics[width=.495\textwidth, trim=30 0 30 0, clip=true]{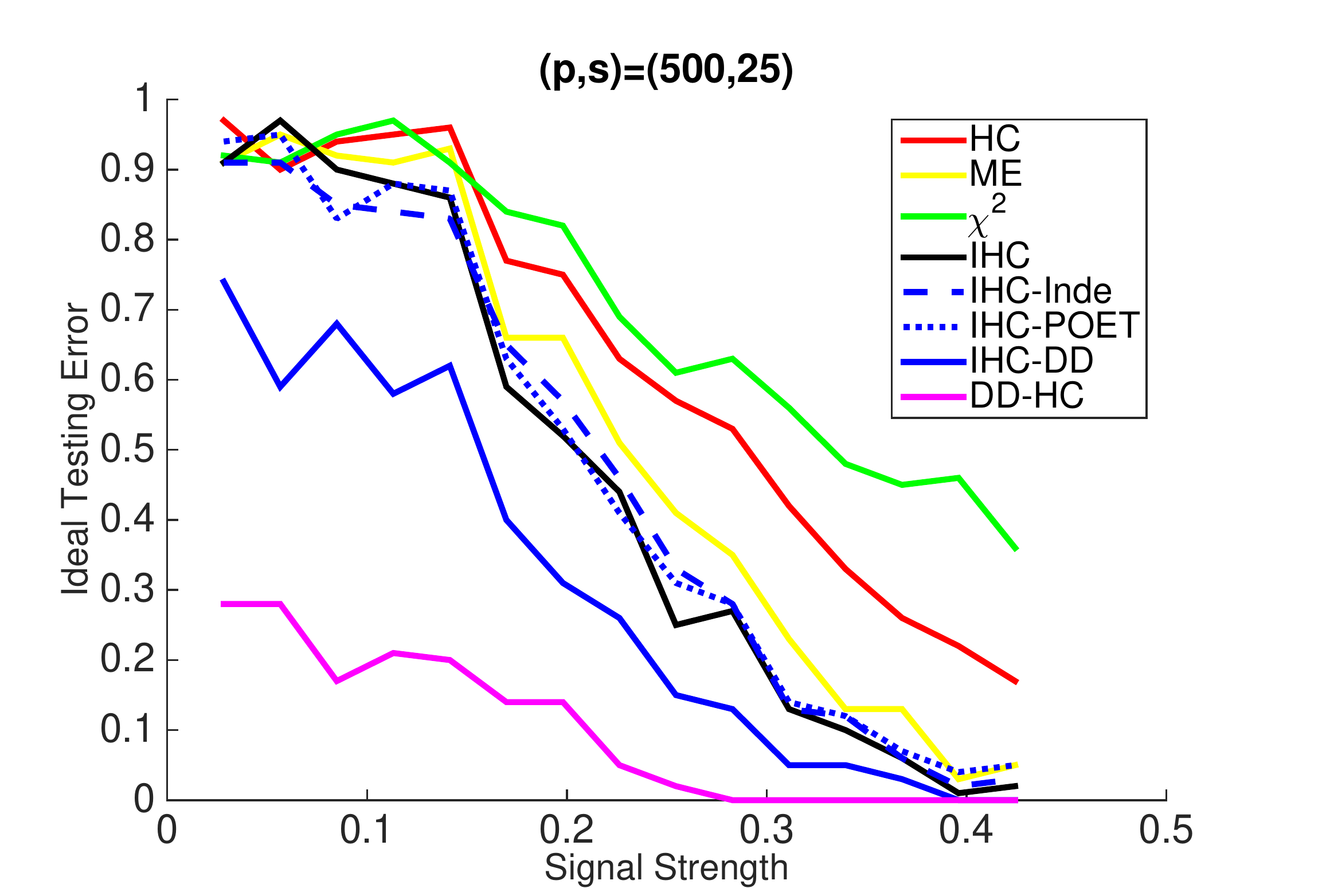}\\
\caption{Ideal testing error (with the best cut-off value of the test statistics).}\label{fig:GlobalTesting}
\end{figure}

We investigate the performance of new tests on extensive simulations.
Given $(n, p,s,\tau)$, let $\mu_j=\tau\cdot 1\{1\leq j\leq s\}$ for $1\leq j\leq p$. Generate the matrix $\bs{\Sigma} = \bs{F}\bs{F}^T + \bs{A}$, where $\bs{F}$ is a $p\times 2$ matrix whose entries are $i.i.d$ drawn from
$\cN(0,1/2)$ and $A_{i,j} = 0.5^{|i-j|}$ for $1\leq i,j\leq
p$. In the null and alternative hypothesis, we generate $X_1,X_2,\ldots,X_n$ $iid$ from ${\cal N}_p(0,\bs{\Sigma})$ and ${\cal N}_p(\mu,\bs{\Sigma})$, respectively. The vector of $z$-scores is $X=\frac{1}{\sqrt{n}}\sum_{i=1}^n X_i$, and $\widehat{\bs{\Sigma}}$ is chosen as the sample covariance matrix of $X_i$'s. For each test, we record the {\it Ideal Testing error}, defined as the sum of type I and type II errors with the optimal cut-off value for this test (computed via $1000$ repetitions). We fix $n=50$ and let $(p, s, \tau)$ take different values, where $s$ controls the sparsity level and $\tau$ controls the signal strength.

We compare our test with the $\chi^2$-test (test statistic: $\|X\|^2$), maximum test (test statistic: $\max_{1\leq j\leq p}|X_j|$), the HC test, and the Innovated HC test ($\widehat{\bs{\Omega}}$ is taken as the generalized inverse of sample covariance matrix). The results are displayed in Figure~\ref{fig:GlobalTesting}. We have several observations. First, the two proposed tests, IHC-DD and DD-HC, significantly outperform the other tests. Especially, the DD-HC test yields  the lowest error in almost all settings. A possible reason is that DD-HC is customized to the factor covariance structure. Second, the IHC-DD test is much better than the IHC test. Since IHC-DD is a variant of IHC by plugging in $\widehat{\bs{\Sigma}}_{ddpca}^{-1}$, for a fair comparison, we include two other variants of IHC by plugging $\widehat{\bs{\Sigma}}_{poet}^{-1}$ and $\widehat{\bs{\Sigma}}_{Ind}^{-1}$, respectively, where $\widehat{\bs{\Sigma}}_{poet}$ is the POET estimator and $\widehat{\bs{\Sigma}}_{Ind}$ is a special case of POET by keeping only the diagonal entries after PCA. Interestingly, these two variants behave similarly as IHC, without much improvement. These observations suggest that the IHC idea is still quite powerful for factor covariance structure, except that we need to plug in a good estimate of $\bs{\Sigma}^{-1}$; they also reconfirm that the DD-PCA covariance estimator does a good job in estimating $\bs{\Sigma}^{-1}$. Last, the $\chi^2$-test, orthodox HC, and maximum test perform unsatisfactorily. Since $\bs{\Sigma}$ is heavily non-sparse, these tests lose power due to not exploring the covariance structure; comparably, the maximum test is less affected.

\section{Optimization for DD-PCA}\label{sec:methodology}

This section studies the optimization problem for DD-PCA. Section~\ref{approximateDDPCA} proposes a  three-block ADMM with provable theoretical guarantees to solve the convex relaxation of DD-PCA. Section~\ref{subsec:iter-alg} proposes an iterative projection algorithm to directly solve the nonconvex optimization of DD-PCA. Section~\ref{alg-discuss} gives a comparison between two approaches.

Before proceeding, we introduce the efficient projection onto $\cSDD_c^+=\cS\cap \cDD_c^+$, the set of ``symmetric $c$-diagonally-dominant'' matrices,
where $\cS$ is the set of symmetric matrices and
$\cDD_c^+$ is the set of $c$-diagonally-dominant matrices with nonnegative
diagonal entries: \be \label{DD} \cDD_c^+ = \{\bs{A} =
(a_{ij})_{p\times p}: a_{jj} \geq c\sum_{i: i\not= j} |a_{ji}|\
\mbox{for\ all\ }j\} \ee It is not difficult to see that both $\cDD_c^+$ and
$\cSDD_c^+$ are closed and convex polyhedral cones. To solve DD-PCA, we shall obtain the (Euclidean) projection
of a matrix $\bs{A}$ onto the convex cone $\cSDD_c^+$ or $\cDD_c^+$, denoted by $\cP_{\cSDD_c^+}(\bs{A})$ or $\cP_{\cDD_c^+}(\bs{A})$. As summarized in Appendix A, the
Mendoza-Raydan-Tarazaga (MRT) algorithm computes the efficient projection $\cP_{\cDD_c^+}(\bs{A})$. Following Theorem 2.1 of \cite*{mendoza-etal-1998}, we have the convergence guarantee that $\bs{X}$ obtained by
MRT Algorithm is the unique projection of $\bs{A}$ onto $\cDD_c^+$. The
computational complexity of MRT Algorithm is $O(p^2\log(p))$.

\subsection{Convex relaxation and ADMM} \label{approximateDDPCA}

This subsection solves the convex relaxation of \eqref{dd-PCA} by replacing nonconvex rank
constraints with convex nuclear norm constraints. To be specific, we consider the convex optimization:
\be\label{convexform2} \min_{(\bs{L},\bs{A})} \
\frac{1}{2}\|\bs{S}-\bs{L}-\bs{A}\|^2_F + \lambda \|\bs{L}\|_*
\quad \text{subject to} \quad \bs{A}\in{\cSDD_c^+}. \ee
where $\lambda$ is a tuning parameter to strike a balance between the approximation error and the low-rank. A large $\lambda$ would
encourage the solution $\hat{\bs{L}}$ to be low-rank, whereas a smaller
$\lambda$ would lead to relatively smaller approximation error but higher rank in $\hat{\bs{L}}$.

We introduce a new variable $\bs{E}$ and rewrite the optimization
problem as follows: \ben \min_{(\bs{L},\bs{A},\bs{E})} \
\frac{1}{2}\|\bs{E}\|^2_F + \lambda \|\bs{L}\|_* + \cI_{\bs{A}\in
\cSDD_c^+} \quad \text{subject to} \quad \bs{L}+\bs{A}+\bs{E}=\bs{S}.
\een
The objective function would be separable in three blocks, subject to an equality constraint. Now, we define the following augmented Lagrange function: \ben
\cL_{\rho}(\bs{L},\bs{A},\bs{E},\bs{\Lambda})   =
\frac{1}{2}\|\bs{E}\|^2_F + \lambda\|\bs{L}\|_{\ast} +
\cI_{\bs{A}\in \cSDD_c^+} +
\frac{\rho}{2}\|\bs{L}+\bs{A}+\bs{E}-\bs{S}\|_F^2 + \langle
\bs{\Lambda}, \bs{L}+\bs{A}+\bs{E}-\bs{S} \rangle \een where
$\bs{\Lambda}$ is the Lagrange multiplier associated with the
equality constraint, and $\rho$ is  a given penalty parameter. The proposed three-block ADMM proceeds as follows till convergence: \bean \bs{L} \
\mbox{step}: \ & \bs{L}^{(t)} = &  \arg\min _{\bs{L}}\ \cL_{\rho}
(\bs{L}, \bs{A}^{(t-1)},\bs{E}^{(t-1)},\bs{\Lambda}^{(t-1)}) \\
 \bs{A} \ \mbox{step}: \ & \bs{A}^{(t)} = &  \arg\min _{\bs{A}}\ \cL_{\rho}
(\bs{L}^{(t)}, \bs{A},\bs{E}^{(t-1)},\bs{\Lambda}^{(t-1)})
\\\bs{E} \ \mbox{step}: \ & \bs{E}^{(t)} = &  \arg\min _{\bs{E}}\ \cL_{\rho} (\bs{L}^{(t)},
\bs{A}^{(t)},\bs{E},\bs{\Lambda}^{(t-1)}) \\
\bs{\Lambda} \ \mbox{step}: \ & \bs{\Lambda}^{(t)} = &
\bs{\Lambda}^{(t-1)} + \rho
(\bs{A}^{(t)}+\bs{L}^{(t)}+\bs{E}^{(t)}-\bs{S}) \eean

Each subproblem can be efficiently solved. In
the $\bs{L}$ step, we solve $\bs{L}^{(t)}$ from  \bean   & \min_{\bs{L}} &
\lambda \|\bs{L}\|_{\ast} +
\frac{\rho}{2}\|\bs{L} + \bs{A}^{(t-1)} + \bs{E}^{(t-1)} -
\bs{S}\|_F^2 + \langle \bs{\Lambda}^{(t-1)}, \bs{L} +
\bs{A}^{(t-1)} + \bs{E}^{(t-1)} - \bs{S} \rangle  \\ \Longleftrightarrow & \min_{\bs{L}} &
 \frac{1}{2}\|\bs{L}+\bs{A}^{(t-1)}+
\bs{E}^{(t-1)}-\bs{S}+\rho^{-1}\bs{\Lambda}^{(t-1)}\|_F^2 +
\rho^{-1}\lambda\|\bs{L}\|_{\ast} \eean
We have $\bs{L}^{(t)} =
\cD_{\rho^{-1}\lambda}\left(\bs{S} -
\bs{A}^{(t-1)}-\bs{E}^{(t-1)}-\rho^{-1}\bs{\Lambda}^{(t-1)}\right)$,
where $\cD_{\tau}$ is the singular value thresholding operator $\cD_{\tau}(\bs{\Omega}) =
\bs{U}s_{\tau}(\bs{D})\bs{V}^T$ for any singular value
decomposition $\bs{\Omega} = \bs{U}\bs{D}\bs{V}^T$, and $s_{\tau}$
denotes the soft-thresholding operator $s_{\tau}(x) =
\mbox{sgn}(x)\max(|x|-\tau,0)$.

In the $\bs{A}$ step, we need to obtain the projection on $\cSDD_c^+$:
\bean \bs{A}^{(t)} & = &
\arg\min_{\bs{A}}\ \cI_{\bs{A}\in \cSDD_c^+} + \frac{\rho}{2}
\left(\|\bs{A} + \bs{L}^{(t)} + \bs{E}^{(t-1)} - \bs{S} +
\rho^{-1}\bs{\Lambda}^{(t-1)}\|_F^2 \right) \\ & = &
\cP_{\cSDD_c^+}\left(\bs{S} - \bs{L}^{(t)} - \bs{E}^{(t-1)} -
\rho^{-1}\bs{\Lambda}^{(t-1)}\right).  \eean
To this end, we follow \cite*{mendoza-etal-1998} to use Dykstra's
alternating projection algorithm between $\cDD_c^+$ and $\cS$. The details of this
alternating projection are included in Appendix A. Alternatively, we may follow the proximal-gradient-based ADMM \citep{ma-etal-2013} to solve the $\bs{A}$ step. See Section 4 of \cite{ma-etal-2013} for more details.

In the $\bs{E}$ step, it is straightforward to solve \bean \bs{E}^{(t)} & = &
\arg\min_{\bs{E}}\ \frac{1}{2}\|\bs{E}\|^2_F + \frac{\rho}{2}
\left(\|\bs{E} + \bs{L}^{(t)} + \bs{A}^{(t)} - \bs{S} +
\rho^{-1}\bs{\Lambda}^{(t-1)}\|_F^2 \right) \\ & = &
\arg\min_{\bs{E}}\ \left\|\bs{E} +
\frac{\rho}{\rho+1}\left(\bs{L}^{(t)} + \bs{A}^{(t)} - \bs{S} +
\rho^{-1}\bs{\Lambda}^{(t-1)}\right) \right\|_F^2 \\ & = &
\frac{\rho}{\rho+1}\left(\bs{S} - \bs{A}^{(t)} - \bs{L}^{(t)} -
\rho^{-1}\bs{\Lambda}^{(t-1)} \right). \eean

Hence, the proposed three-block ADMM can be summarized in
Algorithm~\ref{Convex2}.

\begin{algo}\label{Convex2}
\textbf{ADMM for Solving the Convex Relaxation of DD-PCA}

Given a sample covariance matrix $\bs{S}$, do:
\begin{itemize}
\item Let $\bs{A}^{(0)} = \bs{E}^{(0)} = \bs{\Lambda}^{(0)}=
\bs{0}$.
\item For $t=1,2,\dots$ \begin{itemize} \item
$\bs{L}^{(t)} = \cD_{\rho^{-1}\lambda}\left(\bs{S} -
\bs{A}^{(t-1)}-\bs{E}^{(t-1)}-\rho^{-1}\bs{\Lambda}^{(t-1)}\right)$
where $\cD_{\tau}(\bs{\Omega})$ is the singular value thresholding
operator given by $\cD_{\tau}(\bs{\Omega}) =
\bs{U}s_{\tau}(\bs{D})\bs{V}^T$ for any singular value
decomposition $\bs{\Omega} = \bs{U}\bs{D}\bs{V}^T$, and $s_{\tau}$
denotes the soft-thresholding operator given by $s_{\tau}(x) =
\mbox{sgn}(x)\max(|x|-\tau,0)$.
\item $\bs{A}^{(t)} =
\cP_{\cSDD_c^+}\left(\bs{S} - \bs{L}^{(t)} - \bs{E}^{(t-1)} -
\rho^{-1}\bs{\Lambda}^{(t-1)}\right)$.
\item $\bs{E}^{(t)} =
\frac{\rho}{\rho+1}\left(\bs{S} - \bs{A}^{(t)} - \bs{L}^{(t)} -
\rho^{-1}\bs{\Lambda}^{(t-1)} \right)$.
\item $\bs{\Lambda}^{(t)} =
\bs{\Lambda}^{(t-1)} + \rho
\left(\bs{A}^{(t)}+\bs{L}^{(t)}+\bs{E}^{(t)} - \bs{S}\right)$.
\end{itemize}
\item Stop if the convergence criterion is met.
\end{itemize}
\end{algo}

Although three-block ADMM does not necessarily converge in general \citep{chen-etal-2016}, DD-PCA belongs to a class of regularized least squares decomposition problem. For this class of regularized problems, the global convergence of the proposed three-block ADMM is always guaranteed such that any cluster point of the iterated solutions is an optimal primal and dual pair of DD-PCA (See Theorem 3.2 of \cite{lin2015global}).

\subsection{An iterative projection algorithm}  \label{subsec:iter-alg}

In the sequel, we introduce an iterative projection
algorithm that directly tackles the nonconvex optimization in DD-PCA. The key observation is that we attempt to find a
matrix $\bs{L}^\ast$ in the set $\cL_K = \{\bs{L}:
\text{rank}(\bs{L}) = K\}$ that is closest to the set
$\cM_{\bs{S}} = \{\bs{S}-\bs{A}: \bs{A}\in \cSDD_c^+\}$. Inspired by
\cite{netrapalli-2014}, a natural approach would be to
iteratively project $(\bs{S}-\bs{L})$ onto $\cSDD_c^+$ to update
$\bs{A}$ and then to project $(\bs{S}-\bs{A})$ onto $\cL_K$ to
update $\bs{L}$. To reduce the computational cost, we replace
the projection onto $\cSDD_c^+$ by the projection onto $\cDD_c^+$, followed by symmetrization. Algorithm~\ref{AlPro1} summarizes the details.

\begin{algo}\label{AlPro1} \textbf{Iterative Projection Algorithm for Solving the DD-PCA}

Given a sample covariance matrix $\bs{S}$ and integer $k$, do:
\begin{itemize}
\item Let $\bs{A}^{(0)} = \mathbf{0}$. \item For $t=1,2,\dots$
\begin{itemize} \item $\bs{L}^{(t)} = \cP_{\cL_K}(\bs{S}-\bs{A}^{(t-1)})$ \item $\tilde{\bs{A}}^{(t)} = \cP_{\cDD_c^+}(\bs{S}-\bs{L}^{(t)})$.
\item $\bs{A}^{(t)} = \left(\tilde{\bs{A}}^{(t)} +
(\tilde{\bs{A}}^{(t)})^T\right)/2 $.
\end{itemize}
\item Stop if the convergence criterion is met.
\end{itemize}
\end{algo}

In Algorithm~\ref{AlPro1}, we need to calculate
$\cP_{\cL_K}$ and $\cP_{\cDD_c^+}$. The calculation of
$\cP_{\cDD_c^+}$ is given in Appendix A. The
calculation of $\cP_{\cL_K}$ is given as follows: for any
symmetric matrix $\bs{A}$, we write its eigenvalue decomposition
as $\bs{A}=\bs{Q}\bs{\Lambda}\bs{Q}^T$ where
$\bs{\Lambda}=\diag\{\lambda_1,\lambda_2,\dots,\lambda_p\}$ with
$|\lambda_1| \geq |\lambda_2| \geq \dots \geq |\lambda_p|$. Hence,
the best rank-$K$ approximation is given by $\cP_{\cL_K}(\bs{A})=
\bs{Q}_K\bs{\Lambda}_K \bs{Q}_K^T$ where $\bs{Q}_K$ contains the
first $K$ columns of $\bs{Q}$ and
$\bs{\Lambda}_K=\diag\{\lambda_1,\lambda_2,\dots,\lambda_K\}$.

To use Algorithm~\ref{AlPro1}, we need to estimate the
rank $K$ if it is unknown. A simple estimate of $K$
is to look at the eigenvalues of $\bs{S}$ to pick the $K$ such that
there is a significant gap in magnitude between the first $K$
eigenvalues and the remaining ones. In Section~\ref{sec:simulation}, we
investigate the robustness of the iterative projection algorithm to the
estimation of $K$.

\subsection{Comparing convex and nonconvex approaches} \label{alg-discuss}
The convex
approaches do not require the knowledge of rank $K$ of the low
rank matrix $\bs{L}$, and the global convergence of the proposed ADMM is guaranteed. However, its convergence rate could be slow.
The nonconvex approaches, on the other hand, can be faster in terms of convergence. The
per-iteration cost for Algorithm 2 is
$O(p^2\max\{\log(p),K\})$, compared to $O(p^3)$ for Algorithm 1. But the convergence guarantee of Algorithm 2 is an open question.

The rigorous convergence analysis of the iterative
projection algorithm is difficult due to the non-convexity of the
set $\cL_K$. The existing result (e.g., \cite{drusvyatskiy2015})
proves the local linear convergence of the alternating projections for
two closed sets if the two sets intersect \emph{transversally} at
the converging point. We conjecture that such
condition would hold for most cases in our setting, therefore the convergence would be guaranteed. In practice, our algorithms are stable
and always converge to a valid solution in simulations.

\section{Simulation studies} \label{sec:simulation}

This section investigates several numerical properties of DD-PCA, including the estimation performance, necessity and robustness, and application to covariance matrix estimation.

\paragraph{Experiment 1: Exact DD-PCA.}
We first examine the numerical performance of two-block ADMM (see Appendix B) and the iterative projection algorithm in Algorithm~\ref{AlPro1} to solve the exact DD-PCA. Fixing $(p, K)$,
we first generate a rank-$K$ matrix $\bs{L}=XX^T$ where $X$ is a $p\times k$ matrix whose entries
are i.i.d  drawn from $\cN(0,1/p)$. We then generate a matrix $\bs{A}_0$ with entries
sampled i.i.d from $\cN(0,1/p^2)$ and set $\bs{A} =
\bs{A}_0+\bs{A}_0^T+\bs{D}$, where $\bs{D}$ is a diagonal matrix whose $j$-th diagonal is equal to $\sum_{i:i\neq j}|A_0(j,i)+A_0(i,j)|-2A_0(j,j)$ for $1\leq j\leq p$; it follows that $\bs{A}$ is a diagonally dominant matrix. We then let
$\bs{S} = \bs{L} + \bs{A}$. We consider $p=500,1000,2000$ and fix $K=0.05\cdot p$ for each choice of $p$.

First, for two-block ADMM, we use the solution $(\widehat{\bs{L}},\widehat{\bs{A}})$ after $20$ iterations. Table~\ref{convexexperi} displays the comparison between $(\widehat{\bs{L}},\widehat{\bs{A}})$ and the true $(\bs{L}, \bs{A})$ based on $20$ repetitions. It suggests that $\widehat{\bs{L}}$ always has the same rank as that of $\bs{L}$ and that $\widehat{\bs{L}}$ and $\widehat{\bs{A}}$ are reasonably close to their respective counterparts. Since the ADMM algorithm is an iterative algorithm, $\widehat{\bs{L}}+\widehat{\bs{A}}$ are not exactly equal to $\bs{S}$, but the two matrices are reasonably close after $20$ iterations. The results also suggest that, for a large $p$, more iterations are needed for the convergence of
ADMM.

\begin{table}[h]
\centering
\caption{Performance of the proposed ADMM in Experiment 1.} \vspace{5pt}
\resizebox{0.7\textwidth}{!}{\begin{tabular}{| c | c || c | c | c |
c | } \hline Dimension $p$ & rank($\bs{L}$) & rank($\widehat{\bs{L}}$) &
$\frac{\|\widehat{\bs{L}}+\widehat{\bs{A}} - \bs{S}\|_F}{\|\bs{S}\|_F}$  &
$\frac{\|\widehat{\bs{L}}-\bs{L}\|_F}{\|\bs{L}\|_F}$ & $\frac{\|\widehat{\bs{A}}-\bs{A}\|_F}{\|\bs{A}\|_F}$ \\
\hline 500 & 25 & 25 & 0.008 & 0.011 & 0.045 \\ \hline 1000 & 50 &
50
& 0.010 & 0.008 & 0.034 \\ \hline 2000 & 100 & 100 & 0.013 & 0.006 & 0.026\\
\hline
\end{tabular}} \label{convexexperi}
\end{table}

\begin{figure}[t]
\begin{minipage}[t]{0.5\linewidth}  \centering
\includegraphics[width=\textwidth]{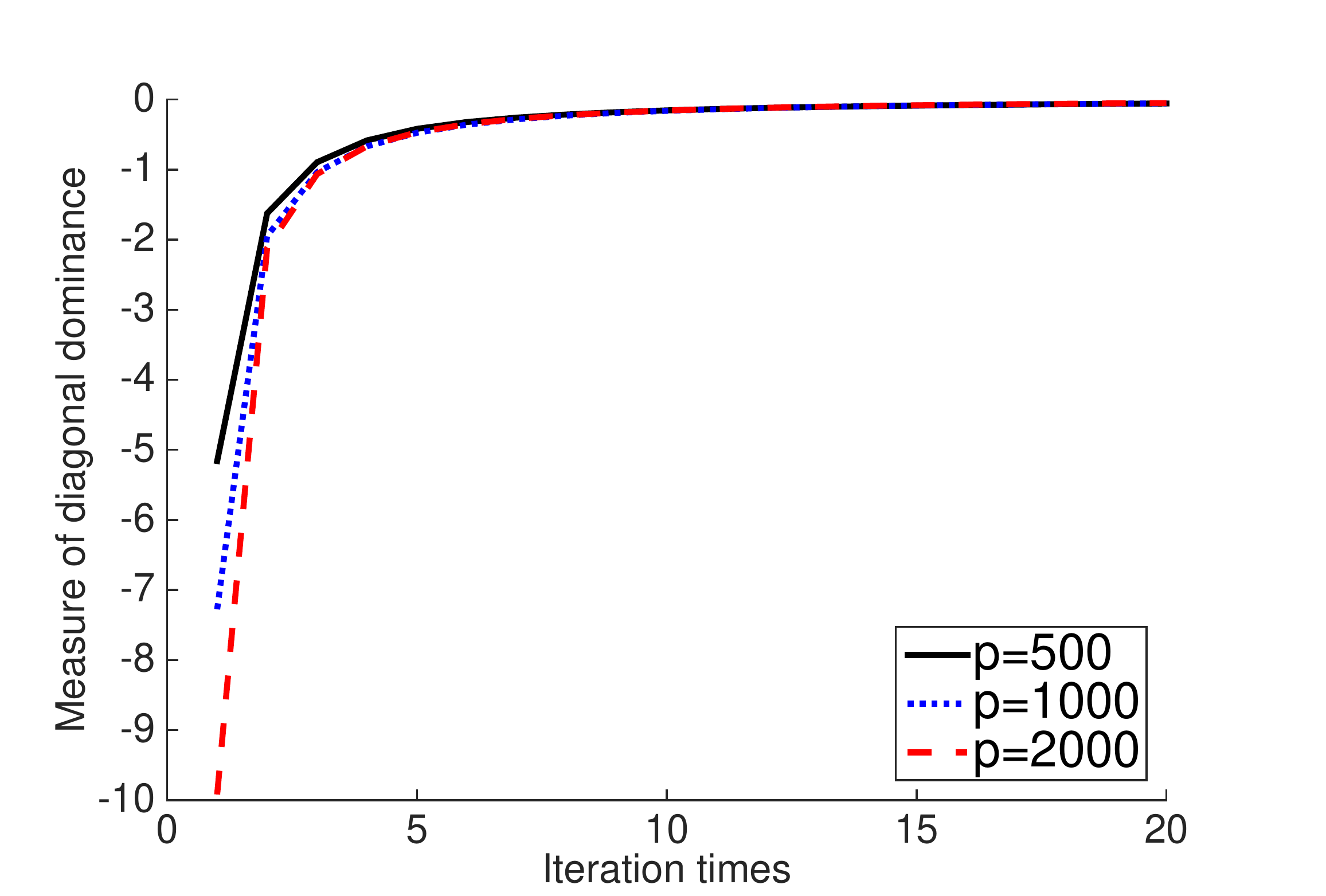} \vspace{-2pt}
\end{minipage}
\begin{minipage}[t]{0.5\linewidth}\centering
\includegraphics[width=\textwidth]{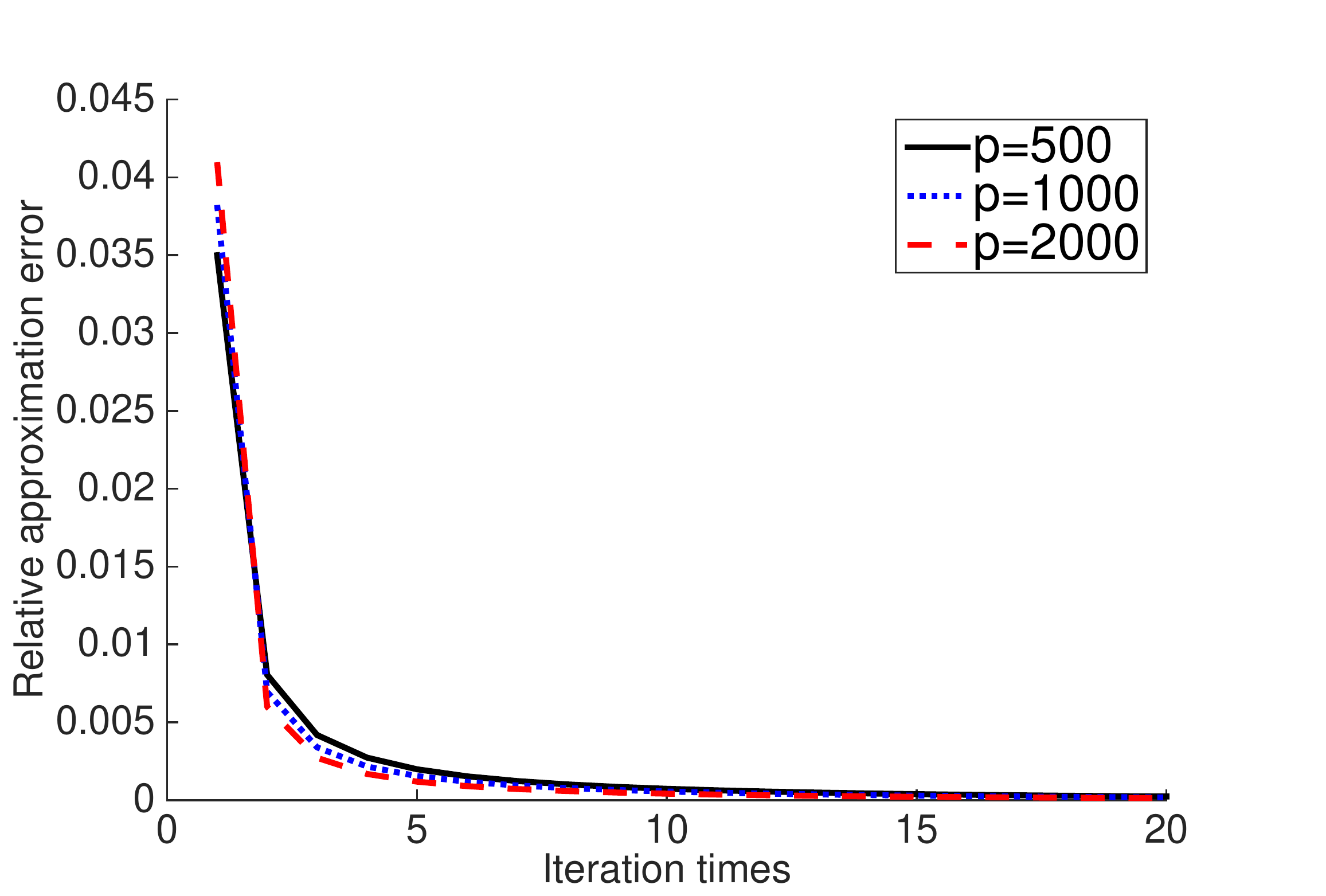} \vspace{-2pt}
\end{minipage}
\vspace{-32pt}
\caption{Performance of Algorithm~\ref{AlPro1} in Experiment 1. The y-axis represents $\zeta(\bs{\Sigma}-\widehat{\bs{L}})$ (left panel) and $\|\widehat{\bs{L}} +\widehat{\bs{A}}
-\bs{S}\|_F/\|\bs{S}\|_F$ (right panel).}\label{iteration}
\end{figure}

Next, we look at Algorithm~\ref{AlPro1}. Instead of fixing the maximum number of iterations, we investigate how the solution $(\widehat{\bs{L}}, \widehat{\bs{A}})$ evolves over iterations.
Note that $\widehat{\bs{L}}$ is guaranteed to have rank $K$ in all iterations, and $\widehat{\bs{A}}$ is equal to the projection of $(\bs{S}-\widehat{\bs{L}})$ into ${\cal DD}^+$ (with symmetrization).  We introduce a quantity to measure the diagonal dominance of $(\bs{S}-\widehat{\bs{L}})$. For any matrix $p\times p$ matrix $\bs{B}$, define
\[
\zeta(\bs{B}) =
\min_{1\leq j\leq p}\Big\{b_{jj}-\sum_{1\leq i\leq p: i\not=j} |b_{ji}|\Big\}.
\]
It measures how close a matrix is to the diagonally dominant cone.
If $\zeta(\bs{\Sigma}-\widehat{\bs{L}})$ continues to increase and eventually gets close to zero, then the algorithm converges. The left panel of Figure~\ref{iteration} shows the evolution of $\zeta(\bs{S}-\widehat{\bs{L}})$ over iterations, and it suggests that the algorithm converges quickly. The right panel of Figure~\ref{iteration} displays the evolution of the relative approximation error $\|\widehat{\bs{L}}
+\widehat{\bs{A}} -\bs{S}\|_F/\|\bs{S}\|_F$, which also decreases quickly over iterations.

\paragraph{Experiment 2: Approximate DD-PCA.}
We investigate the performance of Algorithm~\ref{Convex2} (an ADMM algorithm) and Algorithm~\ref{AlPro1} (an iterative projection algorithm) for approximate DD-PCA.
Fixing $(p,K)$ and $\sigma>0$, we generate a rank $K$ matrix $\bs{L}$ and a diagonally dominant matrix $\bs{A}$ in the same way as in Experiment 1. We then generate a $p\times p$ symmetric matrix $\bs{E}$ whose upper triangular entries are sampled i.i.d from $\cN(0,\sigma^2/p)$. Last, let $\bs{S} = \bs{L} + \bs{A} + \bs{E}$.

First, we study Algorithm~\ref{Convex2}, which is an ADMM algorithm. Fix $\sigma=1$. We consider $p=500,1000,2000$, and set $K=0.05\cdot p$. The tuning parameter in the algorithm is set as $\lambda=3$, and we look at the solution $(\widehat{\bs{L}},\widehat{\bs{A}})$ after 50 iterations. The results are displayed in Table~\ref{convexexperi2}.  For all three settings, the algorithm exactly recovers the true rank of $\bs{L}$, however, the convergence of $(\widehat{\bs{L}},\widehat{\bs{A}})$ is relatively slow. As we shall see below, the performance of Algorithm~\ref{Convex2} is not as good as the iterative projection algorithm---Algorithm~\ref{AlPro1}, but Algorithm~\ref{Convex2} is theoretically more tractable.

\begin{table}[H]
\centering \caption{Performance of Algorithm~\ref{Convex2} in Experiment 2.} \vspace{5pt}
\resizebox{0.7\textwidth}{!}{\begin{tabular}{| c | c || c | c | c |
c | } \hline Dimension $p$ & rank($\bs{L}$) & rank($\widehat{\bs{L}}$) &
$\frac{\|\widehat{\bs{L}}+\widehat{\bs{A}} - \bs{S}\|_F}{\|\bs{S}\|_F}$  &
$\frac{\|\widehat{\bs{L}}-\bs{L}\|_F}{\|\bs{L}\|_F}$ & $\frac{\|\widehat{\bs{A}}-\bs{A}\|_F}{\|\bs{A}\|_F}$ \\
\hline 500 & 25 & 25 & 0.264 & 0.166 & 0.340 \\ \hline 1000 & 50 &
50
& 0.269 & 0.163 & 0.286 \\ \hline 2000 & 100 & 100 & 0.274 & 0.160 & 0.243\\
\hline
\end{tabular}} \label{convexexperi2}
\end{table}

\begin{figure}[ht]
\begin{minipage}[t]{0.5\linewidth}  \centering
\includegraphics[width=\textwidth]{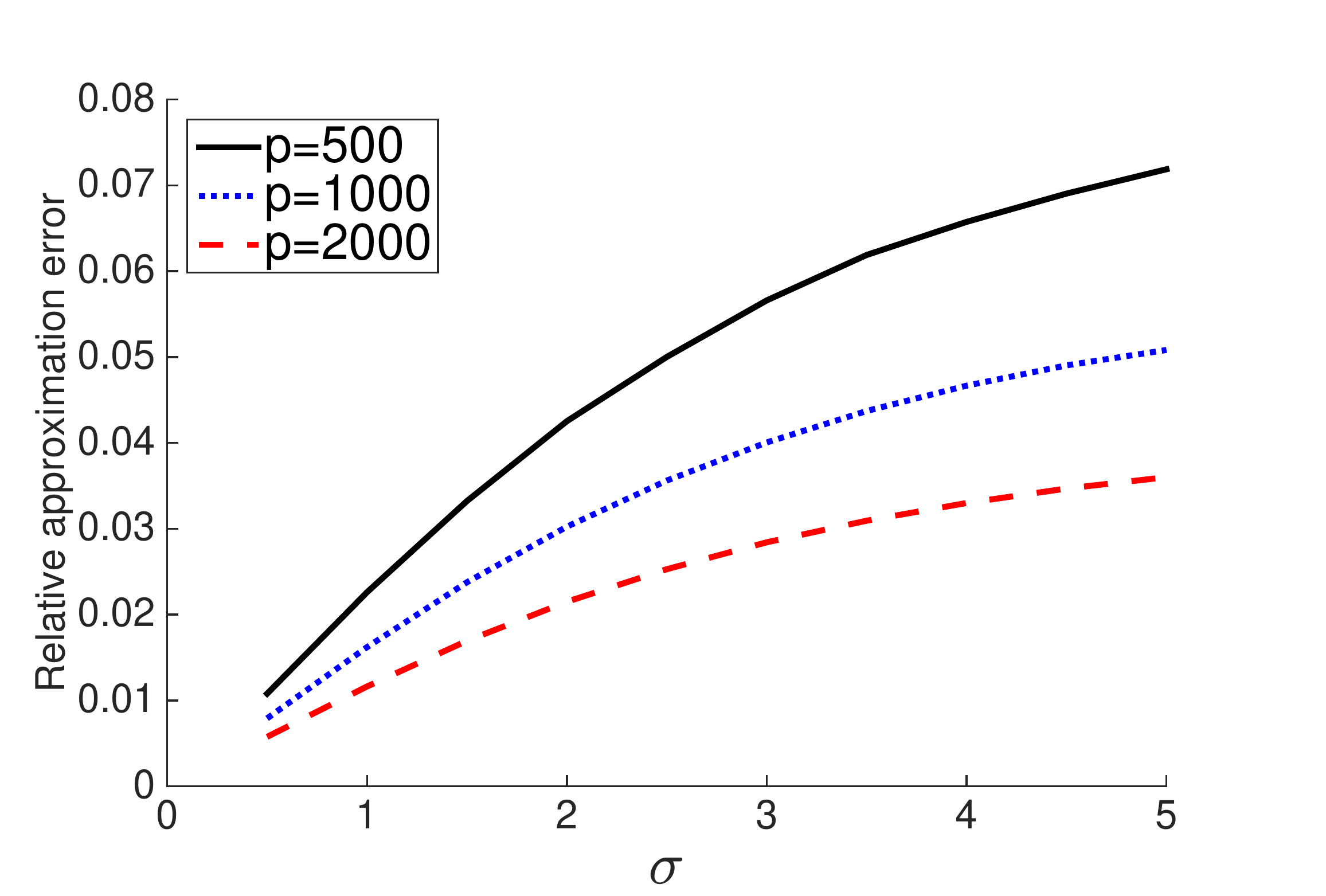} \vspace{-2pt}
\end{minipage}
\begin{minipage}[t]{0.5\linewidth}\centering
\includegraphics[width=\textwidth]{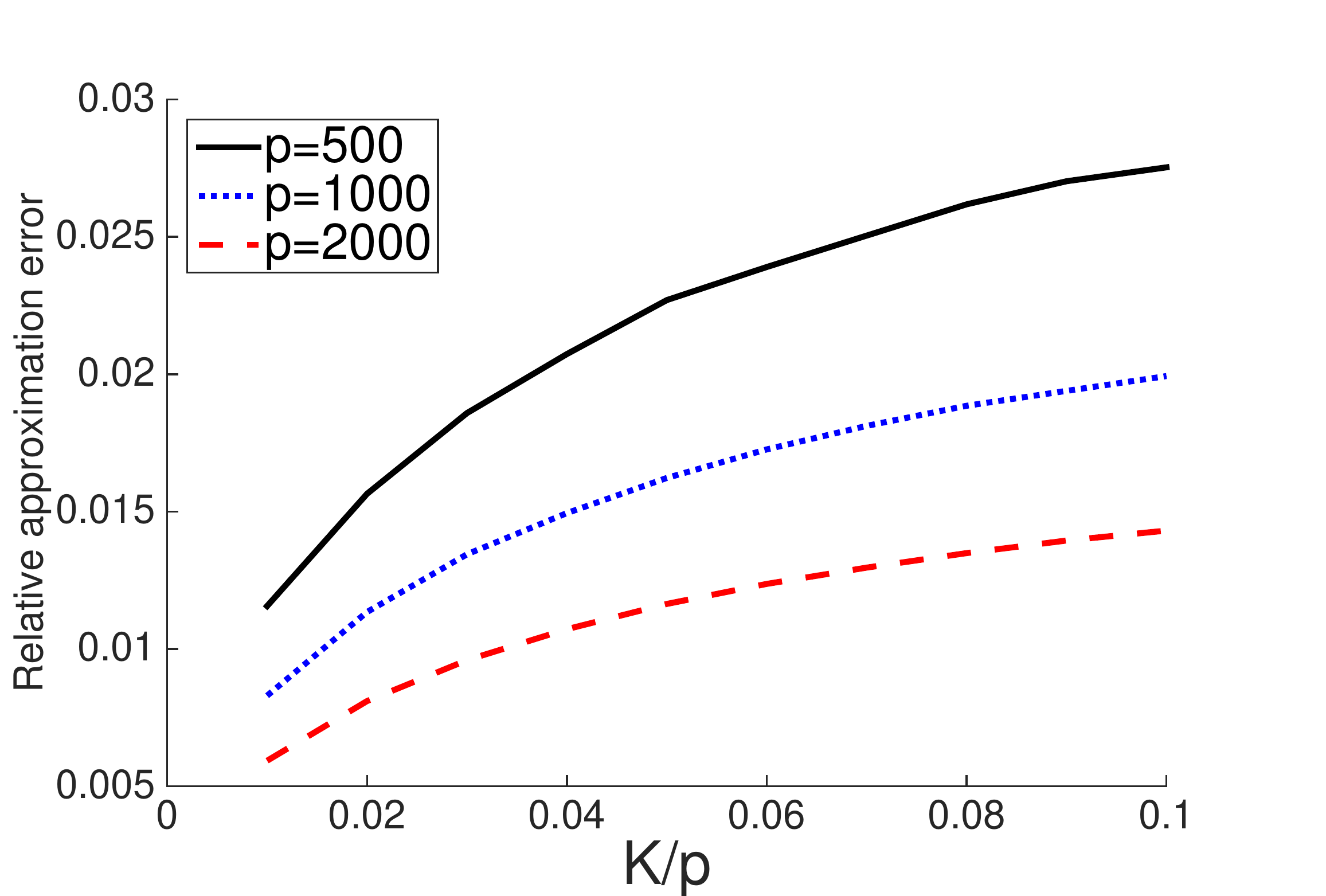} \vspace{-2pt}
\end{minipage}
\vspace{-32pt}
\caption{Performance of Algorithm~\ref{AlPro1} in Experiment 2. The y-axis represents $\|\widehat{\bs{L}} +\widehat{\bs{A}}
-\bs{S}\|_F/\|\bs{S}\|_F$, and the x-axis represents $\sigma$ (left panel) and $K/p$ (right panel), respectively.}\label{iteration2}
\end{figure}

Next, we study Algorithm~\ref{AlPro1}, the iterative projection algorithm. In Experiment 1, we have investigated its performance when $\bs{S}$ has an exact decomposition to the sum of a low-rank matrix and a diagonally dominant matrix. In this experiment, we apply the same algorithm to $\bs{S}$ which does not have such an exact decomposition. We run the algorithm for $20$ iterations and measure the relative approximation error $\|\widehat{\bs{L}} +\widehat{\bs{A}}
-\bs{S}\|_F/\|\bs{S}\|_F$. The results are shown in Figure~\ref{iteration2}. In the left panel, $K/p$ is fixed as $0.05$ and the noise level $\sigma$ varies from $0.5$ to $5$. In the right panel, $\sigma$ is fixed to be $1$ and $K/p$ varies from $0.01$ to $0.1$. For each value of $p$, the relative
approximation error increases, as both $\sigma$ and $K$ increase. For the same values of $\sigma$ and $K/p$, a larger $p$ comes with a
smaller relative approximation error. Furthermore, if we compare the results with those in Table~\ref{convexexperi2}, Algorithm~\ref{AlPro1} has a better practical performance than Algorithm~\ref{Convex2}.

\begin{figure}[t]
\begin{minipage}[t]{0.5\linewidth}  \centering
\includegraphics[width=\textwidth]{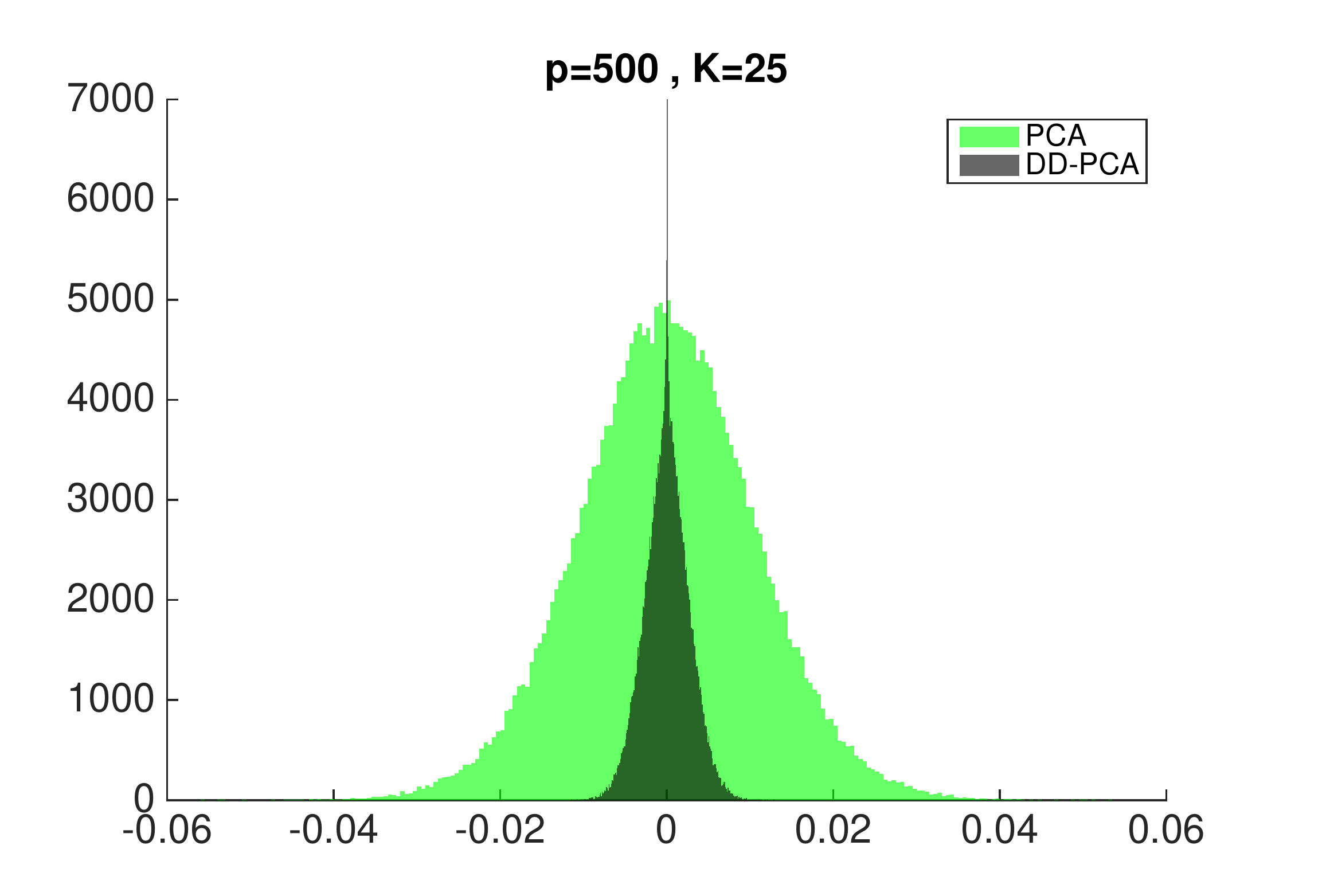} \vspace{-2pt}
\end{minipage}
\begin{minipage}[t]{0.5\linewidth}\centering
\includegraphics[width=\textwidth]{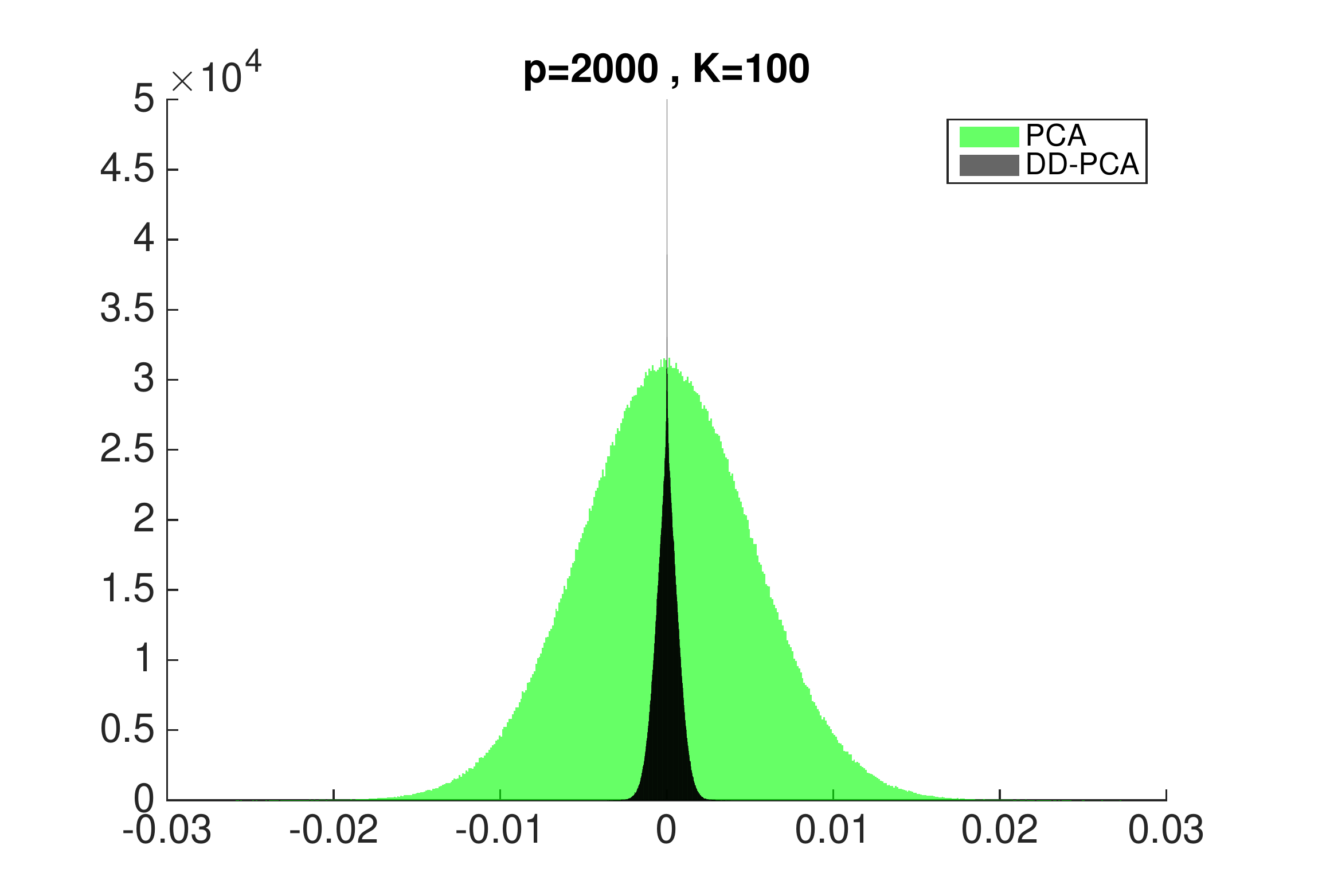} \vspace{-2pt}
\end{minipage}
\vspace{-25pt}
\caption{Comparison of the output
$\bs{A}$ from DD-PCA and from PCA, where the histogram of $\{a_{ij}/[a_{ii}a_{jj}]^{1/2}: 1\leq i\neq j\leq p\}$ is displayed.  
In both panels, the input $\bs{\Sigma}$ is generated as in Experiment 1 in Section~\ref{sec:simulation}.
}\label{Histograms}
\end{figure}

\paragraph{Experiment 3: Necessity of DD-PCA.} If $\bs{\Sigma}$ truly satisfies the assumption of ``low-rank plus diagonal dominance", it is a natural question to know whether one can simply apply PCA and robust PCA \citep{rpca} to get a diagonally dominant $\bs{A}$. Unfortunately, this is often not the case.  Let us consider applying PCA to a $\bs{\Sigma}$ which has the decomposition $\bs{\Sigma}=\bs{L}_0 + \bs{A}_0$ such that $\mathrm{rank}(\bs{L}_0)=K$ and $\bs{A}_0$ is diagonally dominant. Let $\lambda_k$ and $\xi_k$ be the $k$-th eigenvalue and eigenvector, respectively, $1\leq k\leq p$. We construct $\bs{L}=\sum_{k=1}^K\lambda_k\xi_k\xi_k^T$ and $\bs{A}=\bs{\Sigma}-\bs{L}$. We can only hope $\bs{A}$ is diagonally dominant when $\bs{A}$ and $\bs{A}_0$ are entrywise close to each other, or equivalently, when $\|\bs{L}-\bs{L}_0\|_{\max}$ is small ($\|\cdot\|_{\max}$ is the entrywise max norm). However, from the literatures on perturbation analysis of PCA, it requires strong conditions to guarantee that $\|\bs{L}-\bs{L}_0\|_{\max}$ is small \citep{ke2017covariate}. In particular, when $K$ is moderately large, these conditions may be  violated. Similarly, robust PCA cannot produce a diagonally dominant $\bs{A}$ in general. Therefore, it is necessary to develop new algorithms that are specifically designed for DD-PCA. In Figure~\ref{Histograms}, we present a numerical example, where the output $\bs{A}$ from our DD-PCA algorithm is much more ``diagonally dominant" than the $\bs{A}$ constructed from PCA.

\begin{figure}[t]
\begin{minipage}[t]{0.5\linewidth}  \centering
\includegraphics[width=\textwidth]{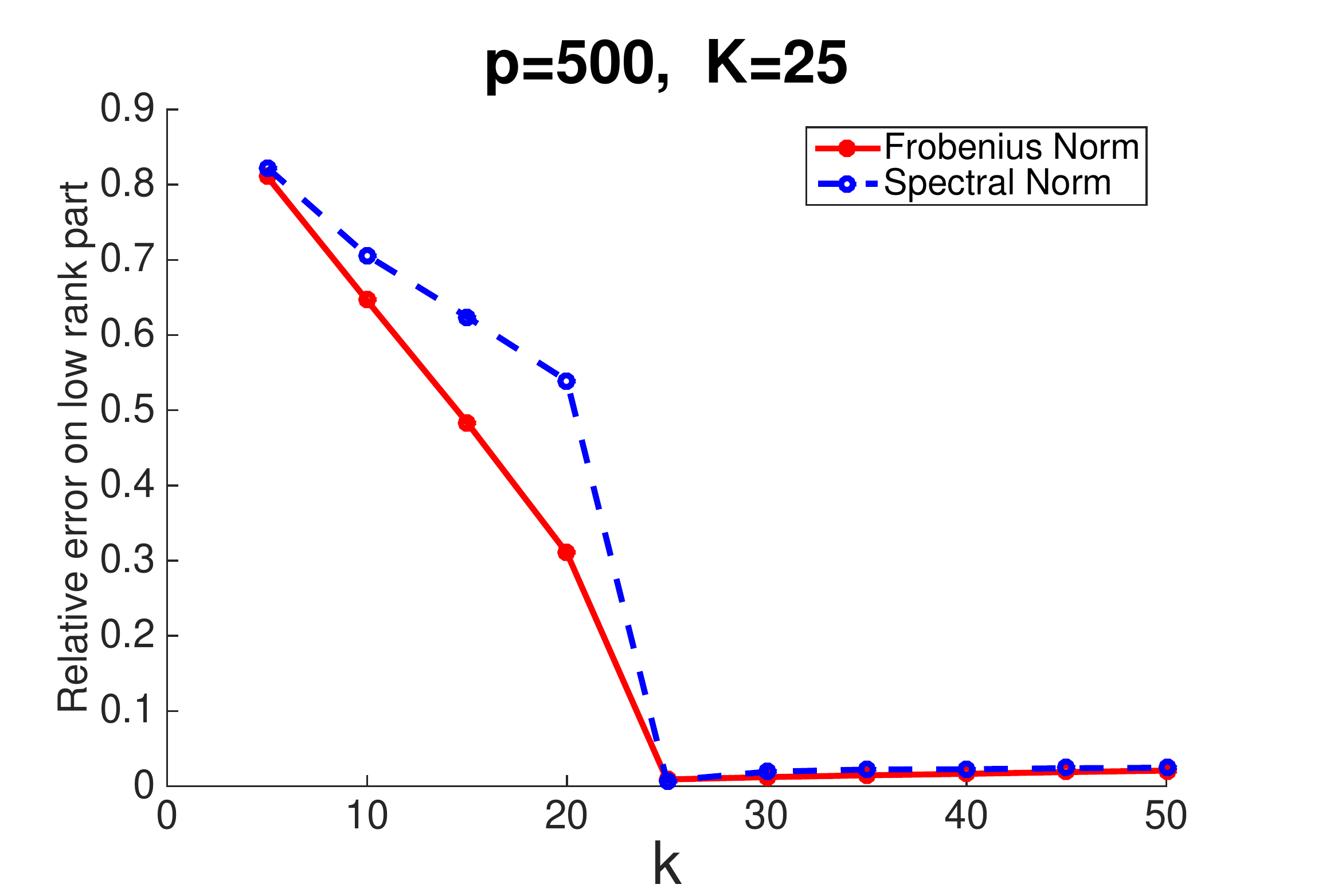} \vspace{-2pt}
\end{minipage}
\begin{minipage}[t]{0.5\linewidth}\centering
\includegraphics[width=\textwidth]{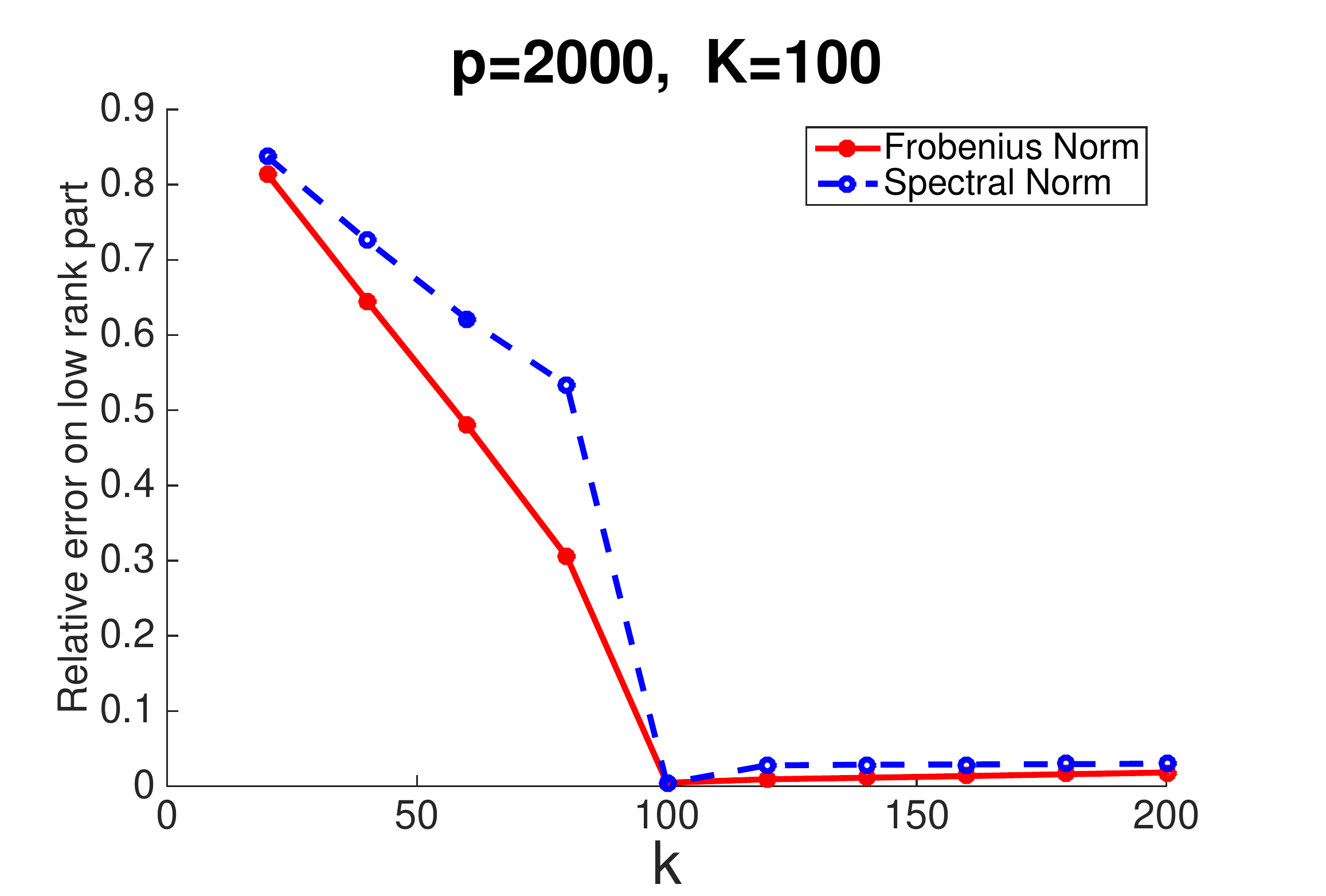} \vspace{-2pt}
\end{minipage}
\vspace{-32pt}
\caption{Robustness of Algorithm~\ref{AlPro1} to a misspecified $K$. The x-axis is the $k$ used in the algorithm, and the y-axis is $\|\widehat{\bs{L}}-\bs{L}\|/\|\bs{L}\|$, where $\|\cdot\|$ is Frobenius norm or spectral norm.}\label{fig:robust}
\end{figure}

\paragraph{Experiment 4: Robustness to the misspecification of
$K$.} We use the same setup as in Experiment 1 and investigate the performance of Algorithm~\ref{AlPro1} with a misspecified $K$.  Consider two settings where $(p,K)=(500,25)$ and $(p,K)=(2000,100)$, respectively. For each setting, we plug $K=k$ in the algorithm and take the solution after ten iterations. Figure~\ref{fig:robust} shows the relative difference between $\widehat{\bs{L}}$ and $\bs{L}$ for various choices of $k$. It suggests that as long as $k\geq K$, the performance of the algorithm is very stable. Hence, in practice, we recommend that the users pick a relatively large $k$ when the true $K$ is hard to estimate.

\paragraph{Experiment 5: Application to covariance matrix estimation}
We expand the numerical study in Section~\ref{sec:CovEst} and investigate the performance of DD-POET on more simulation settings. Given $K=3$ and $p\in\{100,300,500\}$, we generate data in the same way as in the numerical example of Section~\ref{sec:CovEst}. First, we compare the performance of DD-POET and POET. For both methods, we use the true $K=3$. POET has an additional threshold, which we set as the ideal one that minimizes the estimation error (the ideal threshold varies as we change the error measure). The results are contained in Column 6 and Column 10 of Table~\ref{cov_robust}, where, in all settings, DD-POET has a comparable performance as POET with an ideal threshold, and in some settings, DD-POET is even better. The ideal threshold for POET is not practically feasible, and it is unclear how to set the threshold in a data-driven fashion; however, DD-POET is tuning free once $K$ is given. Second, we investigate the performance of DD-POET when we plug in $K=k$ with $k\in\{1,2,\ldots,6\}$; see Table~\ref{cov_robust}. If $k$ is misspecified but $k\geq K$,  the estimation errors remain relatively stable; if $k<K$, the performance deteriorates.  It suggests that an overshooting of $K$ is better than an undershooting. This is consistent with the observations made by \cite{poet}.

\begin{table}[h]
\center \caption{Estimation errors of DD-POET and its robustness to a misspecified $K$.}\vspace{5pt}
\begin{threeparttable}
 \resizebox{0.8\columnwidth}{!}{
\begin{tabular}{c|c|c|c c |c| c c c|c}
\hline \multirow{2}{*}{$(p,K)$}  & \multirow{2}{*}{Target} & \multirow{2}{*}{Norm} & \multicolumn{6}{c|}{$k$} & \multirow{2}{*}{\shortstack[l]{POET$^*$\\ $(k=3)$} }\\
\cline{4-9} & & & 1 & 2 & 3 & 4 & 5 & 6 \\ \hline
\multirow{4}{*}{(100,3)} &
\multirow{2}{*}{$\bs{\Sigma}_u$} &
Frobenius & 48.26 & 27.52 & 3.28 & 3.46 & 3.63 & 3.83 &{\bf 3.24} \\
& & Spectral & 22.51 & 16.80 & {\bf 0.80} & 1.00 & 1.08 & 1.14 & 0.85\\
\cline{2-10} &
\multirow{2}{*}{$\bs{\Sigma}_u^{-1}$} & Frobenius &
9.10 & 7.50 & {\bf 3.02} & 3.17 & 3.34 & 3.56 & 3.65 \\
 & & Spectral & 1.34 & 1.34 & {\bf 0.61} & 0.72 & 0.83 & 0.93 & 0.64 \\ \hline
\multirow{4}{*}{(300,3)} &
\multirow{2}{*}{$\bs{\Sigma}_u$} & Frobenius & 95.00 & 56.96 &
6.22 & 6.23 & 6.26 & 6.32 & {\bf 6.04}\\
 & & Spectral & 32.52 & 26.04 & {\bf 0.82} & 0.86 & 0.90 & 0.93& 0.90\\
\cline{2-10} &
\multirow{2}{*}{$\bs{\Sigma}_u^{-1}$} & Frobenius &
41.50 & 17.33 & {\bf 5.68} & 5.66 & 5.66 & 5.68 & 6.37 \\
& & Spectral & 27.75 & 7.16 & 0.66 & 0.66 & 0.65 & 0.64& {\bf 0.64} \\ \hline
\multirow{4}{*}{(500,3)} &
\multirow{2}{*}{$\bs{\Sigma}_u$} & Frobenius & 126.50 & 75.74 &
8.38 & 8.35 & 8.35 & 8.35 & {\bf 7.99} \\ & &
Spectral & 38.10 & 30.86 & {\bf 0.84} & 0.87 & 0.89 & 0.91& 0.95\\
\cline{2-10} &
\multirow{2}{*}{$\bs{\Sigma}_u^{-1}$} & Frobenius &
25.87 & 18.19 & {\bf 7.66} & 7.61 & 7.57 & 7.55 & 8.26\\
 & & Spectral & 9.23 & 1.76 & 0.69 & 0.68 & 0.68 & 0.67& {\bf 0.64} \\ \hline
\end{tabular}
}\begin{tablenotes}\footnotesize \centering
\item[*] POET is implemented with an ideal threshold.
\end{tablenotes}
\end{threeparttable}
\label{cov_robust}
\end{table}

\section{Discussion}  \label{sec:discussion}

The diagonally dominant matrices have been well studied in linear algebra \citep{feingold1962block,barker1975cones} and optimization \citep{barlow1990computing,mendoza-etal-1998}, motivated by the appealing properties of these matrices for computation. Our work has a very different motivation: We recognize that diagonally dominant (covariance) matrices also have appealing statistical properties, and propose exploring the ``low-rank plus diagonal dominance" covariance structure in data analysis. We demonstrate the benefit of exploring such structure in two statistical problems. For
covariance matrix estimation, we propose DD-PCA as a new estimator. For testing of the global null hypothesis in multiple testing, we propose IHC-DD and DD-HC as two new tests. These new methods have shown encouraging numerical performance in simulations and real applications, especially when the data have a factor covariance structure.

The above methods rely on the availability of algorithms to decompose any given covariance matrix  (approximately) into the sum of a low-rank matrix and a diagonally dominant matrix. To obtain such decomposition is a non-convex optimization. We propose two algorithms --- an ADMM algorithm that solves a convex relaxation, and an iterative projection algorithm that solves the noncovex problem directly. In comparison, the ADMM algorithm is theoretically more tractable, and the iterative projection algorithm shows very appealing numerical performance.

The study here motivates several interesting future directions, such as the uniqueness of the low-rank plus diagonal dominance decomposition, the statistical error of recovering $\bs{L}$ and $\bs{A}$, as well as the convergence of proposed algorithms. We leave to future works.

Our work is related to the literatures of factor models and the literatures of  ``low-rank plus sparse" matrix decomposition. These two lines of works have found wide applications in many areas. Similarly, the use of DD-PCA is not limited to covariance matrix estimation and multiple testing. We expect that DD-PCA will have applications in classification \citep{zhu2005kernel,tong2016survey}, clustering \citep{wang2008variable,clarke2009principles}, dimension reduction \citep{mazhu2013,zou2018selective} and forecasting \citep{fan2017sufficient}.

\newpage

\appendix


\section{Efficient projection onto $\cSDD_c^+$}\label{SDDbackground}

Recall that $\cS$ is the set of symmetric matrices and
$\cDD_c^+$ is the set of $c$-diagonally-dominant matrices with nonnegative
diagonal entries. Now, we present the (Euclidean) projection
of a matrix $\bs{A}$ onto the convex cone $\cSDD_c^+$ or $\cDD_c^+$, denoted by $\cP_{\cSDD_c^+}(\bs{A})$ or $\cP_{\cDD_c^+}(\bs{A})$.

\begin{algo}\label{Dj-proj-full}
\textbf{Mendoza-Raydan-Tarazaga (MRT) Algorithm}

Given a $p\times p$ matrix $\bs{A}$, where the $j$th row of $\bs{A}$
is denoted by $\ba_j$. For $1\leq j\leq p$, the $j$th row of the
projection $\bs{X}$, denoted by $\bx_j$, is given by
\begin{itemize}
\item If $a_{jj}\geq \sum_{l:l\not=j} |a_{jl}|$, then
$\bx_j=\ba_j$.  \item If $-\sum_{l:l\not=j} |a_{jl}| \leq a_{jj} <
0$ and $|a_{jj}|> |a_{jl}|$ for all $l\not=j$, or $a_{jj} <
-\sum_{l:l\not=j} |a_{jl}| $, then $\bx_j=\mathbf{0}$.  \item If
$-\sum_{l:l\not=j} |a_{jl}| \leq a_{jj} < 0$ and $|a_{jj}|\leq
|a_{jl}|$ for some $l\not=j$, or $0\leq a_{jj} < \sum_{l:l\not=j}
|a_{jl}|$,  then $\bx_j$ is generated as follows:
\begin{enumerate}
  \item Sort $|\ba_j|$, excluding $a_{jj}$, in the ascending order, and
  denote the reordered vector as $e$. Note that $e_j=a_{jj}$ and $|e_i| \leq
  |e_l|$ for all $i<l,i\not=j,l\not=j$.
  \item For $m\not=j$, compute $d_m=\sum_{l=m}^p|e_l|\cdot I_{\{j\neq l\}}-e_j$ and $\bar d_m= d_m/(p-m+1)\cdot I_{\{m< j\}}+d_m/(p-m+2)\cdot I_{\{m> j\}}$
  \item Solve $m^\star$ as the smallest integer among $m=1,\ldots,p$ such that $m\neq j$, $|e_m|>0$ and $|e_m|\ge \bar d_m$
  \item Solve $\bx_j=(x_{j1},\dots,x_{jp})$ such that $x_{jj}=a_{jj}+\bar
  d_{m^\ast}$; $x_{ji} = (a_{ji}-\bar
    d_{m^\ast})^+$ if $a_{ji}\geq 0$ for $i\not=j$; $x_{ji} = -(a_{ji}+\bar
    d_{m^\ast})^-$ if $a_{ji}<0$ for $i\not=j$, where $(z)^+ = \max\{z,0\}$ and $(z)^- = -\min\{z,0\}$.
\end{enumerate}
\end{itemize}
\end{algo}

\cite{mendoza-etal-1998} applied Dykstra's
alternating projection algorithm between $\cDD^+$ and $\cS$ to
obtain the projection on $\cSDD^+$. The algorithm is summarized in Algorithm \ref{SDDproj}.

\begin{algo}\label{SDDproj} \textbf{Efficient Projection onto $\cSDD^+$}

Given a $p\times p$ matrix $\bs{A}$,
\begin{itemize}
\item Let $\bs{G}^{(0)} = \bs{A}$ and $\bs{I}^{(0)} = \mathbf{0}$ \item For
$t=1,2,\dots$
\begin{itemize} \item $\bs{G}^{(t)} = \cP_{\cDD^+}\left(\frac{1}{2}(\bs{G}^{(t-1)} + (\bs{G}^{(t-1)})^T) - \bs{I}^{(t-1)}\right)$  \item
$\bs{I}^{(t)} = \bs{G}^{(t)} - \left(\frac{1}{2}(\bs{G}^{(t-1)} + (\bs{G}^{(t-1)})^T)
- \bs{I}^{(t-1)}\right)$
\end{itemize}
\item Stop if the convergence criterion is met.
\end{itemize}

\end{algo}

When $c=1$, the convergence result of Algorithm \ref{SDDproj} can be similarly established as in \cite{boyle1986method} such that the iterated solutions converge in the Frobenius norm to the unique solution of the projection on $\cSDD^+$. More details can be found in \cite{mendoza-etal-1998}. When $c\not=1$, MRT algorithm can't be directly used. In this case, we obtain $\cP_{\cDD_c^+}(\bs{A})$ through Quadratic Programming (QP). The key observation is that the problem can be separated as $p$ independent row-wise projection. For each $1\leq j\leq p$, the $j$th row projection can be written as
\be \label{copt}
\min_{v_1,\dots,v_p} \sum_{i=1}^p ( a_{ji} - v_i )^2 \qquad \text{s.t.} \ v_j \geq c\sum_{i:i\not=j} |v_i|
\ee
and the solution $(v_1,\dots,v_p)$ would be the $j$th row of $\cP_{\cDD_c^+}(\bs{A})$. We can reformulate \eqref{copt} as
\be \label{copt2}
\min_{\delta_1,\dots,\delta_p} \sum_{i=1}^p \delta_i^2 \qquad \text{s.t.} \ a_{jj} - \delta_j  \geq c\sum_{i:i\not=j} | a_{ji} - \delta_i|
\ee
It's easy to see that for $i\not=j$, we should let $\text{sign}(\delta_i) = \text{sign}(a_{ji})$ and $|\delta_i \leq a_{ji}|$, and hence $|a_{ji} - \delta_i| = |a_{ji}| - |\delta_i|$. Without loss of generality, we assume $a_{ji}\geq 0$ for all $i\not=j$ so we can restrict $\delta_i \geq 0$ for all $i\not = j$. Then \eqref{copt2} becomes
\be
\min_{\delta_1,\dots,\delta_p} \sum_{i=1}^p \delta_i^2 \qquad \text{s.t.} \ a_{jj} - \delta_j  \geq c\sum_{i:i\not=j} (a_{ji} - \delta_i), \quad a_{ji}\geq \delta_i \geq 0 \ \text{for all}\ i\not = j
\ee
which is a QP problem and can be solved using standard solver.

\section{Convex relaxation and ADMM for Exact DD-PCA} \label{exactDDPCA}

The exact DD-PCA is difficult to solve
due to the nonconvex rank minimization. Consider the following convex relaxation of the exact DD-PCA:
\be\label{convexform1} \min_{(\bs{L},\bs{A})}
\ \|\bs{L}\|_* \quad \text{subject to} \quad \bs{S} = \bs{L} +
\bs{A}, \ \ \bs{A}\in \mathcal{\cSDD^+}. \ee
where $\|\cdot\|_*$ is the
matrix nuclear norm.

Given the efficient projection onto $\cDD^+$ in Algorithm \ref{Dj-proj-full}, we introduce a new variable $\bs{B}$,
satisfying the equality that $\bs{A}=\bs{B}$, to separate the
symmetric and diagonally-dominant constraints as follows: \ben
\min_{\bs{L},\bs{A}} \|\bs{L}\|_{\ast} + \cI_{\bs{A}\in \cDD^+} +
\cI_{\bs{B}=\bs{B}^T} \quad \text{subject to} \quad \ \bs{S} =
\bs{L}+\bs{A} ,\ \bs{A}-\bs{B}=0 \een
where $\cI_{C}$ is the indicator function which equals to
0 if condition $C$ is satisfied, and equals to infinity otherwise
\citep{boyd2004convex}.

We define the following
augmented Lagrange function: \bean
\cL_{\rho}(\bs{L},\bs{A},\bs{B},\bs{\Lambda}_1,\bs{\Lambda}_2)  &
= & \|\bs{L}\|_{\ast} + \cI_{\bs{A}\in \cDD^+} +
\cI_{\bs{B}=\bs{B}^T}
+ \frac{\rho}{2}(\|\bs{A}-\bs{B}\|_F^2+\|\bs{L}+\bs{A}-\bs{S}\|_F^2) \\
& & + \langle \bs{\Lambda}_1, \bs{A}-\bs{B}\rangle  + \langle
\bs{\Lambda}_2, \bs{L}+\bs{A}-\bs{S} \rangle \eean where
$\bs{\Lambda}_1$ and $\bs{\Lambda}_2$ are the Lagrangian
multipliers associated with the equality constraints, and $\rho$ is  a given penalty parameter. We propose an efficient ADMM to solve the exact DD-PCA from $\cL_{\rho}(\bs{L},\bs{A},\bs{B},\bs{\Lambda}_1,\bs{\Lambda}_2)$,
which proceeds as follows till convergence: \bean \bs{L} \
\mbox{step}: \ & \bs{L}^{(t)} = &  \arg\min _{\bs{L}}\ \cL_{\rho}
(\bs{L},
\bs{A}^{(t-1)},\bs{B},\bs{\Lambda}_1^{(t-1)},\bs{\Lambda}_2^{(t-1)}) \\
\bs{B} \ \mbox{step}: \ & \bs{B}^{(t)} = &  \arg\min _{\bs{B}}\
\cL_{\rho}
(\bs{L}, \bs{A}^{(t-1)},\bs{B},\bs{\Lambda}_1^{(t-1)},\bs{\Lambda}_2^{(t-1)}) \\
\bs{A} \ \mbox{step}: \ & \bs{A}^{(t)} = &  \arg\min _{\bs{A}}\
\cL_{\rho}
(\bs{L}^{(t)}, \bs{A},\bs{B}^{(t)},\bs{\Lambda}_1^{(t-1)},\bs{\Lambda}_2^{(t-1)}) \\
\bs{\Lambda}_1 \ \mbox{step}: \ & \bs{\Lambda}_1^{(t)} = &
\bs{\Lambda}_1^{(t-1)} + \rho (\bs{A}^{(t)}+\bs{L}^{(t)}-\bs{S}) \\
\bs{\Lambda}_2 \ \mbox{step}: \ & \bs{\Lambda}_2^{(t)} = &
\bs{\Lambda}_2^{(t-1)} + \rho(\bs{A}^{(t)} - \bs{B}^{(t)}) \eean

Our proposed ADMM is a two-block ADMM with two blocks
$\{\bs{L},\bs{B}\}$ and $\bs{A}$, and its global convergence is
always guaranteed \citep{boyd2011distributed}. In what follows, we explicitly show how to obtain closed-form
solutions for each subproblem. In the $\bs{L}$ step, we have \bean
\bs{L}^{(t)}  & = & \arg\min_{\bs{L}}\ \|\bs{L}\|_{\ast} +
\frac{\rho}{2}\|\bs{A}^{(t-1)} + \bs{L}-\bs{S}\|_F^2 + \langle
\bs{\Lambda}_1^{(t-1)}, \bs{A}^{(t-1)} + \bs{L} - \bs{S} \rangle
\\ & = & \arg\min_{\bs{L}}\
\frac{1}{2}\|\bs{L}+\bs{A}^{(t-1)}-\bs{S}+\rho^{-1}\bs{\Lambda}_1^{(t-1)}\|_F^2
+ \rho^{-1}\|\bs{L}\|_{\ast} \eean It's easy to show that the
solution is given by $\bs{L}^{(t)} = \cD_{\rho^{-1}}\left(\bs{S} -
\bs{A}^{(t-1)}-\rho^{-1}\bs{\Lambda}_1^{(t-1)}\right)$ where
$\cD_{\tau}(\bs{\Omega})$ is the singular value thresholding
operator given by $\cD_{\tau}(\bs{\Omega}) =
\bs{U}s_{\tau}(\bs{D})\bs{V}^T$ for any singular value
decomposition $\bs{\Omega} = \bs{U}\bs{D}\bs{V}^T$, and $s_{\tau}$
denotes the soft-thresholding operator given by $s_{\tau}(x) =
\mbox{sgn}(x)\max(|x|-\tau,0)$.

In the $\bs{B}$ step, we also have the following closed-form
solution: \bean \bs{B}^{(t)} & = & \arg\min_{\bs{B}}\
\cI_{\bs{B}=\bs{B}^T} + \frac{\rho}{2}\|\bs{A}^{(t-1)} -
\bs{B}\|_F^2 + \langle \bs{\Lambda}_2^{(t-1)},
\bs{A}^{(t-1)}-\bs{B} \rangle
\\ & = & \arg\min_{\bs{B}}\ \cI_{\bs{B}=\bs{B}^T} + \frac{\rho}{2}
\|\rho^{-1}\bs{\Lambda}_2^{(t-1)} + \bs{A}^{(t-1)} - \bs{B}\|_F^2
\\ & = & \frac{1}{2}\left[\left(\bs{A}^{(t-1)} +
\rho^{-1}\bs{\Lambda}_2^{(t-1)}\right) + \left(\bs{A}^{(t-1)} +
\rho^{-1}\bs{\Lambda}_2^{(t-1)}\right)^T \right] \eean

Finally in the $\bs{A}$ step, we have \bean \bs{A}^{(t)} & = &
\arg\min_{\bs{A}}\ \cI_{\bs{A}\in \cDD^+} + \frac{\rho}{2}
\left(\|\bs{A} + \bs{L}^{(t)} - \bs{S} +
\rho^{-1}\bs{\Lambda}_1^{(t-1)}\|_F^2 + \|\bs{A} - \bs{B}^{(t)} +
\rho^{-1}\bs{\Lambda}_2^{(t-1)}\|_F^2 \right) \\ & = &
\arg\min_{\bs{A}}\ \cI_{\bs{A}\in \cDD^+} +  \rho \left\|\bs{A} +
\frac{1}{2}\left(\bs{L}^{(t)} - \bs{S} + \rho^{-1}\bs{\Lambda}_1 -
\bs{B}^{(t)} + \rho^{-1}\bs{\Lambda}_2^{(t-1)}\right)\right\|_F^2
\\ & = & \cP_{\cDD^+}\left(\frac{1}{2}(\bs{S}-\bs{L}^{(t)} +
\bs{B}^{(t)} - \rho^{-1}\bs{\Lambda}_1^{(t-1)} -
\rho^{-1}\bs{\Lambda}_2^{(t-1)})\right) \eean

We summarize our proposed two-block ADMM in Algorithm \ref{Convex1}.

\begin{algo}\label{Convex1}
\textbf{Two-Block ADMM for Solving the Exact DD-PCA}

Given the sample covariance matrix $\bs{S}$, do
\begin{itemize}
\item Let $\bs{A}^{(0)} = \bs{\Lambda}_1^{(0)} =
\bs{\Lambda}_2^{(0)} = \bs{0}$ \item For $t=1,2,\dots$
\begin{itemize} \item $\bs{L}^{(t)} = \cD_{\rho^{-1}}\left(\bs{S} -
\bs{A}^{(t-1)}-\rho^{-1}\bs{\Lambda}_1^{(t-1)}\right)$. \item
$\bs{B}^{(t)} = \frac{1}{2}\left[\left(\bs{A}^{(t-1)} +
\rho^{-1}\bs{\Lambda}_2^{(t-1)}\right) + \left(\bs{A}^{(t-1)} +
\rho^{-1}\bs{\Lambda}_2^{(t-1)}\right)^T \right]$ \item
$\bs{A}^{(t)} = \cP_{\cDD^+}\left(\frac{1}{2}(\bs{S}-\bs{L}^{(t)}
+ \bs{B}^{(t)} - \rho^{-1}\bs{\Lambda}_1^{(t-1)} -
\rho^{-1}\bs{\Lambda}_2^{(t-1)})\right)$ \item
$\bs{\Lambda}_1^{(t)} = \bs{\Lambda}_1^{(t-1)} + \rho
(\bs{A}^{(t)}+\bs{L}^{(t)}-\bs{S})$ \item $\bs{\Lambda}_2^{(t)} =
\bs{\Lambda}_2^{(t-1)} + \rho (\bs{A}^{(t)}-\bs{B}^{(t)})$
\end{itemize}
\item Stop if the convergence criterion is met.
\end{itemize}
\end{algo}

\bibliographystyle{asa}
\bibliography{decorr}

\end{document}